\pdfoutput=1
\documentclass[acmsmall,screen,nonacm]{acmart}
\usepackage{amsthm}
\usepackage{threeparttable}
\usepackage{booktabs}
\usepackage{mathtools}
\usepackage{tabularx}
\usepackage{array}
\usepackage{listings}
\usepackage{stmaryrd}      %
\usepackage{enumitem}
\usepackage{xcolor}
\usepackage{multirow}
\usepackage{tikz}
\usetikzlibrary{arrows.meta, positioning, shapes.geometric, fit, decorations.markings, calc, backgrounds}
\definecolor{frPeach}{HTML}{FBE3D6}
\definecolor{frBlue}{HTML}{CAEEFB}
\definecolor{frGreen}{HTML}{D9F2D0}
\definecolor{frPink}{HTML}{F2CFEE}
\definecolor{frYellow}{HTML}{FFFE9B}
\usepackage{wrapfig}
\usepackage{CJKutf8}
\usepackage{algorithm}
\usepackage{algpseudocode}

\definecolor{codegreen}{rgb}{0,0.6,0}
\definecolor{codegray}{rgb}{0.5,0.5,0.5}

\tikzset{
  posl/.style={gray, font=\scriptsize, inner sep=1pt},
  subtree/.style={draw, regular polygon, regular polygon sides=3, inner sep=1pt, minimum size=7mm},
  cut/.style={postaction={decorate}, decoration={markings,
    mark=at position 0.55 with {
      \draw[thin] (-3.5pt,-3pt) -- (0.5pt,3pt);
      \draw[thin] (-0.5pt,-3pt) -- (3.5pt,3pt);
    }}},
  fragment/.style={fill=#1, rounded corners, inner sep=3pt},
  fragment/.default=frPeach
}
\usepackage{hyperref}
\usepackage{cleveref}
\AddToHook{env/definition/begin}{\crefalias{theorem}{definition}} %

\newcommand{\Real}{\mathrm{Real}}               %
\newcommand{\Sug}{\mathrm{Sug}}                 %

\AtBeginDocument{%
  }

\setcopyright{acmlicensed}
\copyrightyear{2018}
\acmYear{2018}
\acmDOI{XXXXXXX.XXXXXXX}
\acmConference[Conference acronym 'XX]{Make sure to enter the correct
  conference title from your rights confirmation email}{June 03--05,
  2018}{Woodstock, NY}
\acmISBN{978-1-4503-XXXX-X/2018/06}

\begin{document}

\title[Validating Autoformalized Specifications via Contrapuntal Languages]{Validating Autoformalized Specifications\\via Contrapuntal Languages}

\author{Yihan Dai}
\affiliation{%
  \institution{Peking University}
  \city{Beijing}
  \country{China}}
\email{yhdai25@stu.pku.edu.cn}

\author{Xuanlin Liu}
\affiliation{%
  \institution{Nankai University}
  \city{Tianjin}
  \country{China}}
\email{2312289@mail.nankai.edu.cn}

\author{Haoxiang Jia}
\affiliation{%
  \institution{Peking University}
  \city{Beijing}
  \country{China}}
\email{haoxiangjia@stu.pku.edu.cn}

\author{Tianchi Li}
\affiliation{%
  \institution{Peking University}
  \city{Beijing}
  \country{China}}
\email{litianchi@pku.edu.cn}

\author{Xin Zhang}
\affiliation{%
  \institution{Peking University}
  \city{Beijing}
  \country{China}}
\email{xin@pku.edu.cn}

\author{Sergey Mechtaev}
\authornote{Corresponding author.}
\affiliation{%
  \institution{Peking University}
  \city{Beijing}
  \country{China}}
\email{mechtaev@pku.edu.cn}

\renewcommand{\shortauthors}{Dai et al.}

\begin{abstract}
Formal specifications can make AI-generated code mechanically checkable. Because user intent is typically expressed in natural language rather than formal logic, specifications must first be derived from it, e.g., via LLM-based autoformalization. Yet autoformalized specifications are themselves untrusted, so a key challenge is validating whether they capture user intent without relying on a costly oracle such as human review. In this oracle-free setting, a natural strategy is to vote among model outputs, but because model errors are often correlated, agreement among outputs may reflect shared mistakes rather than correctness.
Our key insight is that these errors can be decorrelated through representational diversity. We achieve this diversity with \emph{contrapuntal languages}: pairs of task-specific formal languages, each designed to express the intended specification, yet admitting no structure-preserving translation between them. An LLM formalizes the intent independently in each language, and we accept a specification only if a solver proves the two formalizations equivalent. Because the model must discover separately how to express the intent in each language's constructs, agreement is less likely to reflect a correlated error than agreement within a single language. To construct such languages, we present a synthesis algorithm that reuses fragments of sampled seed specifications as syntactic sugar over a base specification language.
We implement the approach for first-order logic specifications and evaluate it on HumanEval+ and LiveCodeBench using GPT-5.4 mini and Qwen3.6-Flash with and without thinking. Over the strongest baseline among direct generation, plurality voting, and round-trip verification, our method increases the proportion of accepted specifications that are correct, i.e., complete and sound, by 12.0\% on HumanEval+ and 42.9\% on LiveCodeBench. The accepted specifications judged complete are more effective at detecting bugs: they reject 98.6\% of buggy program variants on HumanEval+ and 96.2\% on LiveCodeBench, compared with 92.8\% and 78.3\% for the strongest baseline. A case study shows how disagreement between contrapuntal languages can surface ambiguity in a natural-language task. 
\end{abstract}



\newcommand{\yihan}[1]{{\color{red}\textbf{(Yihan:} #1\textbf{)}}}
\newcommand{\xuanlin}[1]{{\color{red}\textbf{(Xuanlin:} #1\textbf{)}}}

\maketitle

\section{Introduction}
\label{sec:intro}

As large language models (LLMs) generate code faster than humans can review it~\cite{mozannar2024reading,duma2026these}, the central challenge is shifting from generating code to determining whether that code faithfully captures the user's intent. A promising answer is \emph{intent autoformalization}: automatically translating natural-language intent into formal specifications against which generated code can be mechanically verified. Yet this only relocates the trust problem: if the specification is itself produced by a fallible autoformalization process, does it capture the user's intent expressed in natural language? Prior studies on HumanEval tasks show that generating behaviorally correct postconditions and contracts from natural-language intent remains challenging for LLMs~\cite{endres2024largelanguagemodelstransform,richter2025postconditionslargelanguagemodels,prasetya2026talk}. For example, even for Python contracts on HumanEval, the setting most familiar to LLMs, only 39.5\% of GPT-5-generated contracts that raised no false alarms on the reference implementation rejected every evaluated mutant~\cite{richter2025postconditionslargelanguagemodels}. However, validating generated specifications typically requires costly human review, making oracle-free validation a central bottleneck for reliable AI-assisted software development~\cite{lahiri2026intent}.

How can we validate whether autoformalized specifications accurately capture the user's intent without an external oracle? A straightforward approach is voting. Plurality voting~\cite{hansen1990neural, NEURIPS2024_6034a661} partitions the specifications into equivalence classes and selects the largest, i.e., the most frequently occurring semantics. However, LLM-generated artifacts often exhibit correlated errors~\cite{kim2025correlated}; when multiple candidates share the same mistake, plurality voting can reinforce the dominant error rather than identify the correct answer~\cite{tumer1996error}. Also, plurality voting treats each candidate as an indivisible vote, ignoring its internal structure: a candidate need not be correct as a whole for some of its components to be useful. Clause2Inv~\cite{cao2025clause2inv}, for example, shows that clauses extracted from invalid invariant candidates can contribute to a valid invariant. Yet reusing such fragments naively would steer the model back toward the same formalizations, and thus the same errors. The challenge is therefore to reuse useful fragments while inducing sufficiently different formalization paths.

To address these challenges, we build on the insight that representational diversity can break the correlation between errors. Inspired by linguistic determinism---the hypothesis that language shapes thought~\cite{whorf1956language}---we introduce \emph{contrapuntal languages}: pairs of task-specific formal languages, each designed to express the intended specification, yet admitting no structure-preserving translation between them. An LLM formalizes the intent independently in each language, and we accept a specification only if a solver proves the two formalizations equivalent. Because the model must discover separately how to express the intent in each language's constructs, cross-language agreement is less likely to reflect a correlated error than agreement within a single language.

To construct such languages, we present a synthesis algorithm that builds both languages as syntactic sugar over a base specification language. We first sample seed specifications for the task from the LLM; the algorithm then cuts each seed into fragments and turns them into language constructs, using a solver to choose cuts that carve every seed into differently shaped trees in the two languages. Since the intended specification is unknown, the seeds stand in for it: we call a language pair \emph{effectively contrapuntal} for a seed set when every seed can be represented in both languages, but no AST-level structure-preserving translation relates its representations across them. This design also addresses the challenge of reusing fragments: seed fragments let the model reuse useful components, even from incorrect seeds, while the structural incompatibility of the two languages still forces separate formalization paths and thus decorrelates errors.

We implement the approach for first-order logic specifications in SMT-LIB and evaluate it on HumanEval+~\cite{liu2023your} and LiveCodeBench~\cite{jain2025livecodebench} using GPT-5.4 mini and Qwen3.6-Flash with and without thinking. We compare it against direct generation, plurality voting, and round-trip verification, judging a specification complete if it accepts reference input--output pairs and sound if it rejects the outputs of buggy program variants. Over the strongest baseline, our method increases the proportion of accepted specifications that are both complete and sound by 12.0\% on HumanEval+ and 42.9\% on LiveCodeBench, and the accepted specifications judged complete reject 98.6\% and 96.2\% of buggy program variants, respectively, compared with 92.8\% and 78.3\% for the strongest baseline. An ablation shows that these gains depend on enforcing contrapuntality rather than merely using two languages, and a case study shows that disagreement between contrapuntal languages can surface ambiguity in a task description.

In summary, the paper makes the following contributions:
\begin{itemize}
    \item A new approach that automatically validates autoformalized specifications without an oracle through \emph{contrapuntal languages}.
    \item An algorithm to automatically synthesize contrapuntal languages with an SMT solver.
    \item An evaluation on HumanEval+ and LiveCodeBench showing that contrapuntal languages accept a higher proportion of complete and sound specifications than three baselines, and that the accepted specifications judged complete detect more bugs.
\end{itemize}

All code and data are available at \url{https://figshare.com/s/be97e0291c8ce608bcd4}.

\section{Motivating Example}
\label{sec:motivating_example}

To illustrate the difficulty of validating autoformalized specifications
without access to any intent oracle, we use HumanEval/135 as a running
example. The task asks \texttt{can\_arrange(arr: list[int])} to return the largest index $i$ such that
$\mathit{arr}[i] < \mathit{arr}[i-1]$, or $-1$ if no such index exists.

\begin{figure}[t]
    \centering
    \includegraphics[width=1\linewidth]{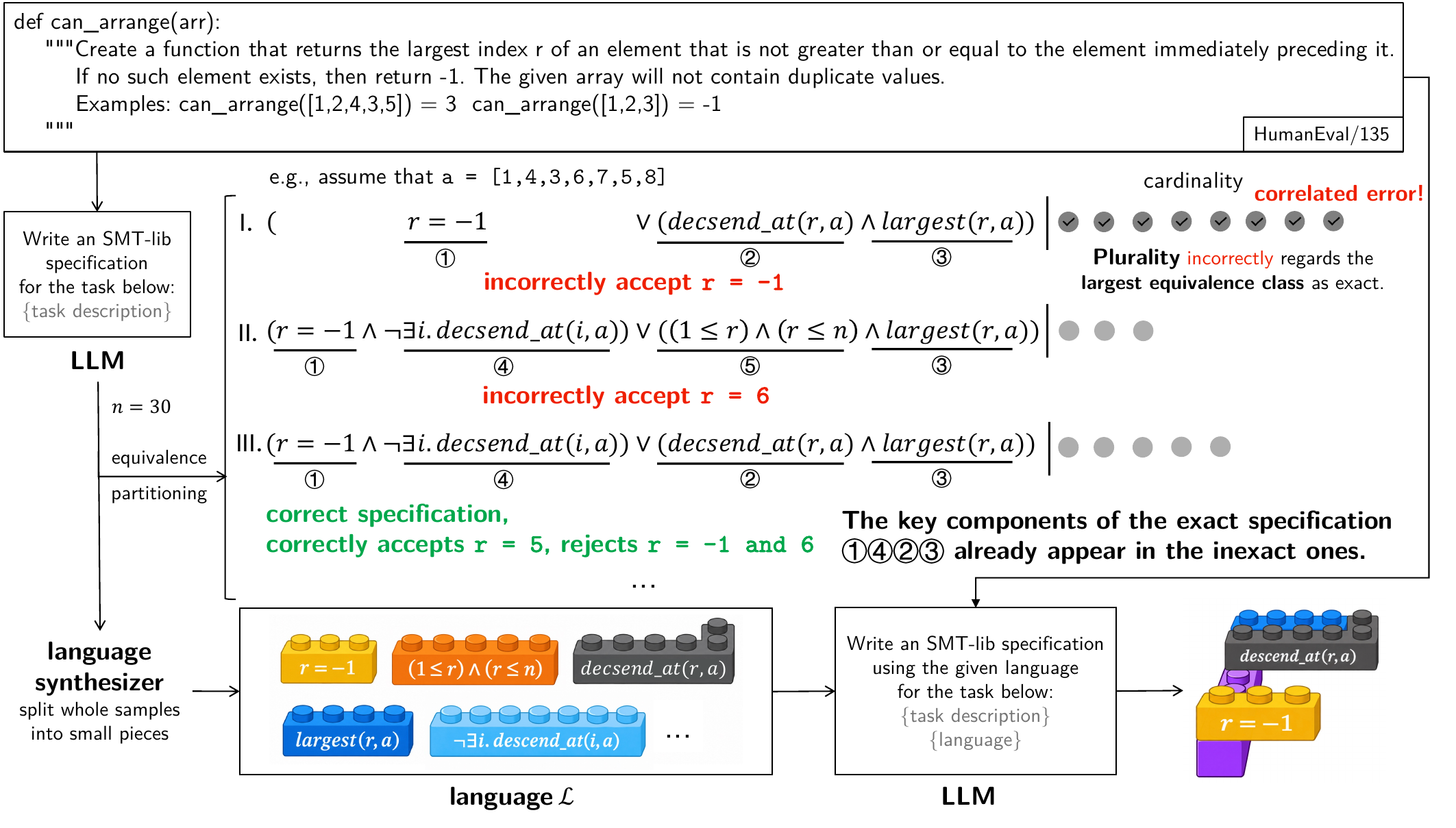}
    \setlength{\abovecaptionskip}{-8pt}
        \caption{Plurality voting on HumanEval/135 with $n=30$ samples. A correlated error (I, 8/30) outnumbers the correct specification (III, 5/30), so Plurality accepts it. Yet the key components of III already appear in the incorrect I and II, motivating language synthesis from such fragments. $\mathit{descend\_at}(i,a) \triangleq (1 \le i < |a|) \land (a[i] < a[i-1])$; $\mathit{largest}(r,a) \triangleq \forall i.\,(\mathit{descend\_at}(i,a) \Rightarrow i \le r)$.} 
    \label{fig:comparison}
\end{figure}

When using Qwen3.6-Flash (thinking=on) to autoformalize this task into an SMT-LIB specification, it has a low success rate: the estimated probability of sampling a correct specification (formula III in ~\Cref{fig:comparison}) is only 0.17, while the dominant incorrect one (the first formula in ~\Cref{fig:comparison}) has the probability $ 0.27\ (8/30) > 0.17\ (5/30)$. This incorrect specification does accept the intended output on every observed positive example, but it also permits $r=-1$ when such valid index $r$ exists. It is therefore complete but unsound.

Plurality voting selects the most frequent semantic equivalence class among all sampled specifications. It can be ineffective because LLMs tend to produce correlated errors---a phenomenon observed not only across samples from the same model but even across different models~\cite{kim2025correlated}. In challenging cases like HumanEval/135, plurality voting amplifies rather than mitigates mistakes: as shown in \Cref{fig:comparison}, it selects the dominant incorrect class over the correct one. Plurality voting also underuses the information in the candidate specifications, using them only to count equivalence classes. Yet a candidate need not be correct as a whole for some of its components to contribute to a correct specification: in \Cref{fig:comparison}, the key components of the correct specification already appear in the incorrect ones. We therefore split candidates into small pieces that, like the building bricks in \Cref{fig:comparison}, form a small, task-specific vocabulary. We call this process ``synthesizing a language from seed formalizations''.

\begin{figure}[t]
    \centering
    \includegraphics[width=1\linewidth]{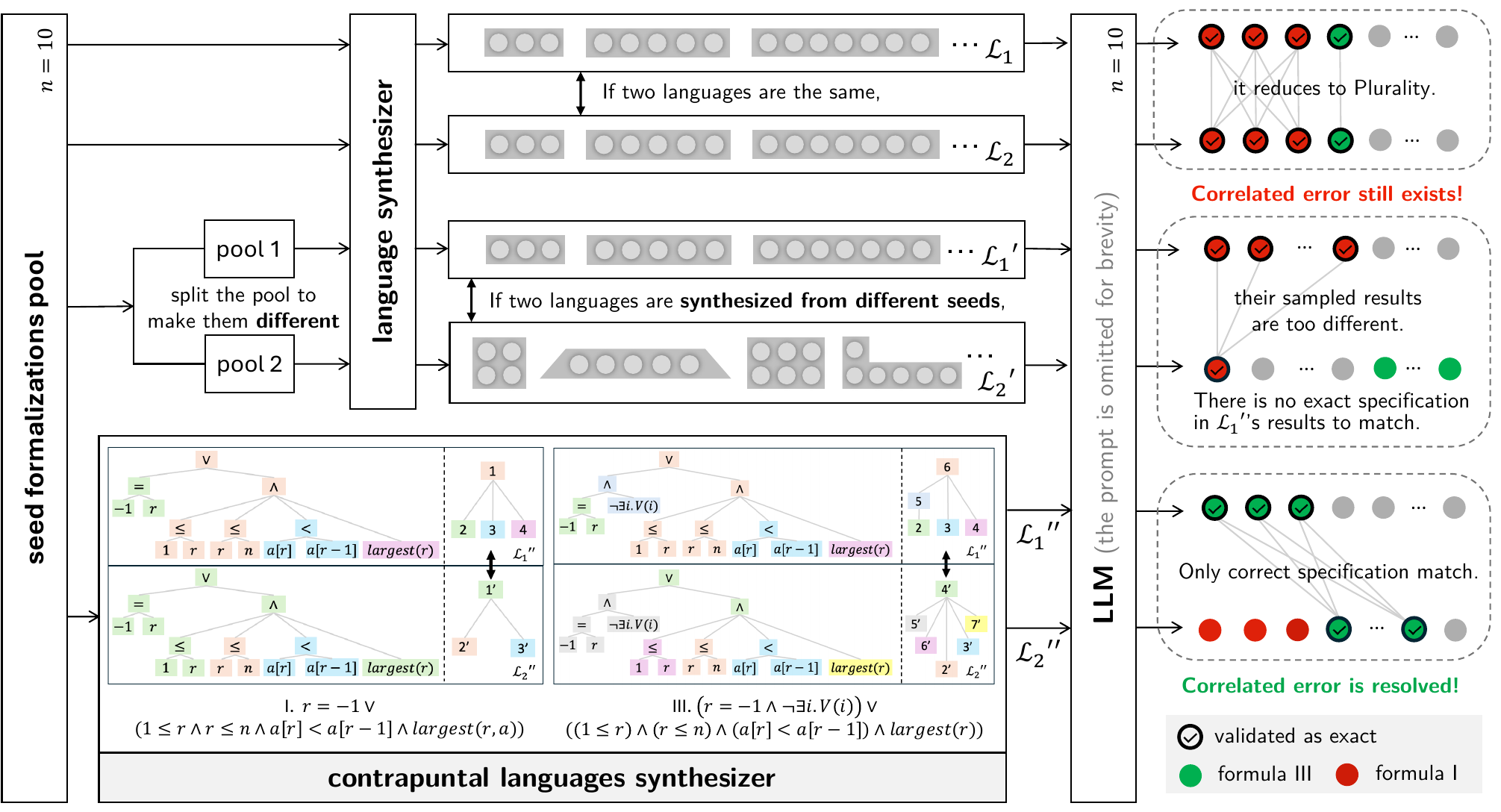}
    \setlength{\abovecaptionskip}{-6pt}
    \caption{Autoformalizing in identical languages (top) reduces to plurality voting; in languages from split seeds (middle), outputs diverge yet share the error. Contrapuntal languages (bottom) both express the correct formalization but are structurally incompatible, so the LLM does not repeat its mistakes.}
    \label{fig:show_tree}
\end{figure}

Autoformalization in a single language still suffers from correlated errors. To decorrelate them, we exploit linguistic determinism: we force the model to ``think'' differently by formalizing the intent in two different languages synthesized from the seeds. \Cref{fig:show_tree} illustrates why these languages must be chosen carefully. If both are synthesized from the same seeds (top row), they may be identical, and cross-language validation reduces to plurality voting, preserving the correlated error. A natural remedy is to randomly split the seeds into two subsets and synthesize one language from each (middle row), yielding $\mathcal{L}_1'$ and $\mathcal{L}_2'$. This split-seed strategy, however, fails in two ways. First, it makes the languages too different: each exposes the LLM to different fragments, so their outputs diverge. Formalizations under $\mathcal{L}_1'$ contain no correct specification, whereas those under $\mathcal{L}_2'$ contain seven, so no correct specification can be accepted. Second, it does not make them different enough: splitting the seeds gives no control over correlated errors, and the incorrect formalization under $\mathcal{L}_2'$ is equivalent to seven under $\mathcal{L}_1'$, so the error is accepted anyway. Our ablation study (\Cref{sec:rq2}) shows that these failures are systematic rather than specific to this example.

These two failures suggest that the languages require a delicate balance: similar enough that correct formalizations can be matched across them, yet different enough that errors do not recur in both. We capture this balance with a property we call \emph{contrapuntality} (formalized in \Cref{def:contrapuntal}), which combines two requirements. \emph{Adequacy} requires that both languages can express the intended specification; unlike split-seed languages, they thus offer comparable expressive power, so a correct formalization in one language has a counterpart in the other. \emph{Non-isomorphism} requires that no structure-preserving translation relates formalizations across the two languages; the model therefore cannot reuse the structure of a formalization in one language when writing it in the other, making it less likely to repeat the same error in both. Here, structure refers to the shape of abstract syntax trees (ASTs): languages that differ only in vocabulary produce ASTs of the same shape for the same formula, so validating across them reduces to plurality voting.

With contrapuntal languages (bottom row of \Cref{fig:show_tree}), the example is handled correctly. On the one hand, the incorrect seeds are cut into different fragments and thus have different AST structures in the two languages, making it difficult for the same error to recur in both. On the other hand, since both languages are synthesized from all seeds, each can extract critical structures from the correct seeds---specifically, structures that prevent the LLM from unconditionally accepting $r = -1$. For example, $\mathcal{L}_1''$ obtains construct~5 and $\mathcal{L}_2''$ obtains construct~5$'$. Formalizations in both languages use these structures, form equivalent pairs, and are therefore accepted.

\noindent
\begin{minipage}[t]{0.52\textwidth}
\hspace*{1em}The distinction matters when the validated specification is used to check generated code. Consider this buggy implementation of \texttt{can\_arrange}, whose backward scan stops before index~1:
\end{minipage}
\hspace{2mm}
\begin{minipage}[t]{0.44\textwidth}
\vspace{-3mm}
\begin{lstlisting}[language=Python,escapechar=|]
def can_arrange(arr):
    for i in range(len(arr) - 1, |{\setlength{\fboxsep}{0.8pt}\colorbox{red!25}{1}}|, -1):
        if not (arr[i] >= arr[i - 1]):
            return i
    return -1
\end{lstlisting}
\end{minipage}

On input $\mathit{arr}=[6,0,1,2,3]$, the mutant returns $r=-1$, whereas the intended result is $r=1$ because $\mathit{arr}[1] < \mathit{arr}[0]$. Formula~I, selected by plurality voting, admits the incorrect input--output pair $(\mathit{arr},-1)$, so the mutant survives the specification check. In contrast, the specifications accepted by our method realize Formula~III, which permits $r=-1$ only when no decreasing adjacent pair exists. Because $\mathit{descend\_at}(1,\mathit{arr})$ holds, Formula~III is unsatisfiable for $r=-1$, so it rejects the incorrect behavior and kills the mutant. Thus, more reliable specification validation translates directly into stronger bug detection. Beyond bug detection, disagreement between contrapuntal languages can also surface ambiguity in the task description, as our case study shows (\Cref{sec:rq4}).

\section{Background \& Notation}
\label{sec:background}

Presenting our approach requires writing formulas in formal languages while also reasoning about these languages. We therefore distinguish the \emph{object languages}, in which formulas are written, from the \emph{meta-language}, in which we reason about them. Greek letters are placeholders, called \emph{meta-variables}: $\varphi, \psi, \chi$ stand for arbitrary formulas and $\pi, \rho$ for arbitrary positions. All other symbols stand for themselves, with two exceptions: $x$ stands for an arbitrary object variable, and the names we define for particular formulas and variables, such as $\mathrm{Def}^i$ and $c^i_\pi$ in \Cref{sec:synth-shape}, stand for what they name. Thus, $p(v) \land q(v)$ is a concrete formula, whereas $\varphi \land \psi$ denotes the conjunction of the formulas that $\varphi$ and $\psi$ stand for. We also distinguish three kinds of equality: $=$ is meta-level identity, which for formulas means identity of their syntax trees; $\doteq$ is the equality symbol of the object language; and $\equiv$ is semantic equivalence (defined below). For example, $\varphi \land \varphi \neq \varphi$, yet $\varphi \land \varphi \equiv \varphi$.

Our approach targets specifications written in first-order logic, the common input language of SMT solvers, standardized by SMT-LIB~\cite{barrett2010smt}; this lets us check the equivalence of formalizations automatically. These formulas form the base language $\mathcal{L}_0$: they are built as usual from atoms, object equality $\doteq$, the connectives $\lnot, \land, \lor, \rightarrow, \leftrightarrow$, the constants $\top$ and $\bot$, and the quantifiers $\forall$ and $\exists$. We interpret formulas over fixed domains, such as the integers and arrays, in which function and predicate symbols have fixed meanings. The \emph{denotation} $\llbracket \varphi \rrbracket$ of a formula $\varphi$ is the set of variable assignments that make $\varphi$ true. For example, $\llbracket r \doteq -1 \lor r \geq 1 \rrbracket$ consists of the assignments that map $r$ to $-1$ or to a positive integer. For a specification whose free variables are the inputs and outputs of a program, the denotation is thus the set of input--output pairs that the specification admits. Formulas $\varphi$ and $\psi$ are \emph{semantically equivalent}, written $\varphi \equiv \psi$, iff $\llbracket \varphi \rrbracket = \llbracket \psi \rrbracket$.

\begin{definition}[Language]
\label{def:language}
A \emph{language} $\mathcal{L}$ consists of (i)~an alphabet $\Lambda$ of symbols, each with an arity $\mathit{ar}(\alpha) \in \mathbb{N}$; (ii)~a set of \emph{$\mathcal{L}$-formulas}, written $\psi \in \mathcal{L}$: finite trees in which every node labeled $\alpha \in \Lambda$ has $\mathit{ar}(\alpha)$ children, possibly subject to further well-formedness constraints; and (iii)~a \emph{semantics} assigning to every $\mathcal{L}$-formula $\psi$ a denotation $\llbracket \psi \rrbracket$, a set of variable assignments, so that semantic equivalence applies across languages. We write $\Lambda_i$ for the alphabet of a language $\mathcal{L}_i$.
\end{definition}

Our approach compares formulas by the shapes of their syntax trees, so we need notation for addressing the parts of a formula.\label{sec:bg-trees} A \emph{position} is a finite string of positive integers: the empty string $\varepsilon$ addresses the root, and $\pi i$ addresses the $i$-th child of the node at $\pi$. For a formula $\varphi$, $\mathrm{Pos}(\varphi)$ is its set of positions, $\varphi(\pi)$ is the \emph{label} of the node at $\pi$, and $\varphi|_\pi$ is the subtree rooted there. Bound variables are folded into quantifier labels: the root of $\forall x.\, \varphi$ is a unary node labeled $\forall x$, so that positions reflect logical structure only. We write $\pi \sqsubseteq \pi'$ if $\pi$ is a prefix of $\pi'$, that is, if $\pi'$ lies in the subtree at $\pi$, and $\pi \sqsubset \pi'$ if moreover $\pi \neq \pi'$.

We synthesize new languages as syntactic sugar over the base language $\mathcal{L}_0$. Each construct of a new language abbreviates a fixed pattern of $\mathcal{L}_0$-formulas, so the new languages change how specifications are written but not what they mean.\label{sec:bg-sugar}

\begin{definition}[Sugared language]
\label{def:sugar}
A \emph{sugared language} $\mathcal{L}$ has an alphabet $\Lambda$ of \emph{constructs} and, for every construct $\alpha \in \Lambda$ of arity $k$, a \emph{definition} $D_\alpha$. The definition is a finite tree over the symbols of $\mathcal{L}_0$ and fresh \emph{holes} $\square_1, \dots, \square_k$, in which each hole occurs at least once and at least one node is not a hole, such that filling the holes with $\mathcal{L}_0$-formulas $\chi_1, \dots, \chi_k$ yields an $\mathcal{L}_0$-formula $D_\alpha[\chi_1, \dots, \chi_k]$. The $\mathcal{L}$-formulas are the trees over $\Lambda$, and \emph{desugaring} replaces every construct by its definition:
\[
D(\chi) \;\triangleq\; D_\alpha[D(\chi|_1), \dots, D(\chi|_k)], \quad\text{where } \alpha = \chi(\varepsilon) \text{ and } k = \mathit{ar}(\alpha).
\]
An $\mathcal{L}$-formula denotes what its desugaring denotes, $\llbracket \chi \rrbracket \triangleq \llbracket D(\chi) \rrbracket$; in particular, $\chi \equiv D(\chi)$.
\end{definition}

We say that $\chi$ \emph{desugars to} an $\mathcal{L}_0$-formula $\psi$ iff $D(\chi) = \psi$ without variable capture: every bound variable occurrence in $\psi$ comes from the same node of $\chi$ as its quantifier. The \emph{sugarings} of $\psi$ in $\mathcal{L}$ form the set $\Sug_{\mathcal{L}}(\psi) \triangleq \{\chi \in \mathcal{L} \mid \chi \text{ desugars to } \psi\}$; In our supplementary materials, we illustrate these notions on Formula~I of \Cref{fig:comparison}.

Finally, we define correctness for \emph{autoformalization}, the translation of a natural-language specification into a formula.\label{sec:bg-autoformalization} We write $\eta$ for such an informal specification and $\mathit{NL}$ for the set of all of them. We do not analyze their syntax; we only assume that each $\eta$ has a denotation $\llbracket \eta \rrbracket$: the set of variable assignments, such as input--output pairs, that the user intends (for ambiguous text, the reading the user has in mind). We write $\psi \equiv \eta$ iff $\llbracket \psi \rrbracket = \llbracket \eta \rrbracket$. Unlike equivalence between formulas, this cannot be checked automatically, because $\llbracket \eta \rrbracket$ is known only to the user.

A formula $\psi$ is \emph{sound} for $\eta$ iff $\llbracket \psi \rrbracket \subseteq \llbracket \eta \rrbracket$: every behavior it admits is intended. It is \emph{complete} for $\eta$ iff $\llbracket \eta \rrbracket \subseteq \llbracket \psi \rrbracket$: every intended behavior is admitted. A \emph{correct formalization} of $\eta$ is both sound and complete, that is, $\psi \equiv \eta$. For example, Formula~I in \Cref{fig:comparison} is complete but unsound: it admits every intended output, but also $r = -1$ on the input $[6,0,1,2,3]$, whose intended output is $1$. Conversely, $\mathit{descend\_at}(r,a) \land \mathit{largest}(r,a)$ is sound but incomplete: it rejects the intended output $-1$ on inputs without a descent. Formula~III is both sound and complete, and thus a correct formalization.

\begin{definition}[Realization set]
\label{def:realization}
The \emph{realization set} of an informal specification $\eta$ in a language $\mathcal{L}$ collects the correct formalizations of $\eta$ in $\mathcal{L}$:
\[
\Real_{\mathcal{L}}(\eta) \;\triangleq\; \{\psi \in \mathcal{L} \mid \psi \equiv \eta\}.
\]
The language $\mathcal{L}$ is \emph{adequate} for $\eta$ iff $\Real_{\mathcal{L}}(\eta) \neq \emptyset$.
\end{definition}

\section{Contrapuntal Languages}
\label{sec:contrapuntal}

As \Cref{sec:motivating_example} showed, cross-checking formalizations in two languages is useful only if the languages are chosen well. First, both languages must be able to express the specification correctly; we call this \emph{adequacy}. Second, the languages must differ enough that the LLM is unlikely to make the same mistake in both; we call this \emph{non-isomorphism}. We refer to a pair of languages with both properties as \emph{contrapuntal}, after musical counterpoint, where independent melodies combine into one harmony. In this section, we define these properties formally and then relax them to a version that can be checked automatically; \Cref{sec:synthesis} shows how to synthesize languages that satisfy it.

\begin{definition}[Structure-preserving translation]
\label{def:scheme}
A \emph{structure-preserving translation} from $\chi_1 \in \mathcal{L}_1$ to $\chi_2 \in \mathcal{L}_2$ is a one-to-one map $\tau$ from the labels of $\chi_1$ to the labels of $\chi_2$ that turns $\chi_1$ into $\chi_2$ by relabeling: $\mathrm{Pos}(\chi_1) = \mathrm{Pos}(\chi_2)$ and $\chi_2(\pi) = \tau(\chi_1(\pi))$ for every $\pi \in \mathrm{Pos}(\chi_1)$.
\end{definition}

\begin{definition}[Contrapuntal languages]
\label{def:contrapuntal}
Languages $\mathcal{L}_1$ and $\mathcal{L}_2$ are \emph{contrapuntal} for an informal specification $\eta \in \mathit{NL}$ iff
\begin{enumerate}[label=(\roman*)]
\item \emph{adequacy}: both $\mathcal{L}_1$ and $\mathcal{L}_2$ are adequate for $\eta$; and
\item \emph{non-isomorphism}: no formula in $\Real_{\mathcal{L}_1}(\eta)$ has a structure-preserving translation to a formula in $\Real_{\mathcal{L}_2}(\eta)$.
\end{enumerate}
\end{definition}

Definition~\ref{def:contrapuntal} captures our intuition, but it cannot be checked algorithmically, for two reasons. First, both clauses refer to realization sets, and hence to $\llbracket \eta \rrbracket$, which is known only to the user. Second, even with known semantics, non-isomorphism requires proving that synthesizing a translation is \emph{unrealizable}, which is undecidable in general~\cite{hu2019proving}.

We make the problem tractable by weakening it in two ways. First, in place of the inaccessible semantics of $\eta$, we use a finite set of \emph{seed formalizations}: candidate formalizations of $\eta$ sampled from the LLM in the base language. The seeds stand for the candidate meanings of $\eta$---the intended formalization together with its plausible near-miss variants. Second, we restrict the synthesized languages to syntactic sugar over the base language, and constrain only the sugarings of the seeds.

\begin{definition}[Effective contrapuntality]
\label{def:effective}
Sugared languages $\mathcal{L}_1$ and $\mathcal{L}_2$ are \emph{effectively contrapuntal} for a finite, nonempty set $\Psi$ of $\mathcal{L}_0$-formulas iff, for every $\psi \in \Psi$:
\begin{enumerate}[label=(\roman*)]
\item \emph{adequacy}: $\Sug_{\mathcal{L}_1}(\psi) \neq \emptyset$ and $\Sug_{\mathcal{L}_2}(\psi) \neq \emptyset$; and
\item \emph{non-isomorphism}: no formula in $\Sug_{\mathcal{L}_1}(\psi)$ has a structure-preserving translation to a formula in $\Sug_{\mathcal{L}_2}(\psi)$.
\end{enumerate}
\end{definition}

The two definitions differ only in replacing the realization sets $\Real_{\mathcal{L}_i}(\eta)$, which contain all formulas equivalent to the intent, with the sugaring sets $\Sug_{\mathcal{L}_i}(\psi)$, which contain only the formulas that reproduce the syntax of the seed. Unlike realization sets, sugaring sets are finite and computable for a finite alphabet: every construct covers at least one node of $\psi$, so a sugaring of $\psi$ is never larger than $\psi$ itself. Whether a translation maps one given formula to another is decided by comparing the two trees position by position. Effective contrapuntality is therefore decidable.

\begin{example}
\label{ex:effective}
Consider two seeds that differ in a single connective, $\psi_\rightarrow = p(v) \rightarrow q(v)$ and $\psi_\land = p(v) \land q(v)$, and let $\Psi = \{\psi_\rightarrow, \psi_\land\}$. Consider also two sugared candidate languages for $\Psi$: the constructs of $\mathcal{L}_1$ fuse the connective with the antecedent, whereas those of $\mathcal{L}_2$ mirror the base symbols one by one:
\[
\begin{array}{@{}l@{\quad}l@{\qquad}l@{\qquad}l@{\qquad}l@{}}
\mathcal{L}_1\colon & D_{\alpha_\rightarrow} = p(v) \rightarrow \square_1, & D_{\alpha_\land} = p(v) \land \square_1, & D_{\alpha_q} = q(v); \\
\mathcal{L}_2\colon & D_{\beta_\rightarrow} = \square_1 \rightarrow \square_2, & D_{\beta_\land} = \square_1 \land \square_2, & D_{\beta_p} = p(v), & D_{\beta_q} = q(v).
\end{array}
\]

\noindent
\begin{minipage}[t]{0.55\textwidth}
Each seed has one sugaring per language: $\psi_\rightarrow$ has $\alpha_\rightarrow(\alpha_q)$ and $\beta_\rightarrow(\beta_p, \beta_q)$, and $\psi_\land$ has $\alpha_\land(\alpha_q)$ and $\beta_\land(\beta_p, \beta_q)$. The $\mathcal{L}_1$-sugarings have two nodes and the $\mathcal{L}_2$-sugarings three, so no translation relates them, and $\mathcal{L}_1$ and $\mathcal{L}_2$ are effectively contrapuntal for $\Psi$.

\hspace*{1em}This balance between $\mathcal{L}_1$ and $\mathcal{L}_2$ is delicate. Extend $\mathcal{L}_2$ to $\mathcal{L}_2'$ with a construct that fuses the connective with the \emph{consequent}, $D_{\gamma_\rightarrow} = \square_1 \rightarrow q(v)$. Now $\psi_\rightarrow$ has a second sugaring, $\gamma_\rightarrow(\beta_p)$, and the relabeling $\alpha_\rightarrow \mapsto \gamma_\rightarrow$, $\alpha_q \mapsto \beta_p$ translates $\alpha_\rightarrow(\alpha_q)$ into it (Figure~\ref{fig:effective-example}), so $\mathcal{L}_1$ and $\mathcal{L}_2'$ are no longer effectively contrapuntal. Note that this translation pairs $\alpha_q$, which stands for $q(v)$, with $\beta_p$, which stands for $p(v)$: a translation only relabels the tree, regardless of what the constructs stand for.
\end{minipage}\hfill
\begin{minipage}[t]{0.41\textwidth}
\vspace{0pt}
\centering
\begin{tikzpicture}[
  level distance=7mm,
  level 1/.style={sibling distance=16mm},
  every node/.style={inner sep=2pt}
]
\begin{scope}
\node (i1) {$\rightarrow$}
  child { node (p1) {$p$} child { node (pv1) {$v$} } }
  child { node (q1) {$q$} child { node (qv1) {$v$} } };
\begin{pgfonlayer}{background}
\node[fragment=frPeach, fit=(i1)(p1)(pv1)] (f1) {};
\node[fragment=frBlue, fit=(q1)(qv1)] (g1) {};
\end{pgfonlayer}
\node[posl, anchor=east] at (f1.west) {$\alpha_\rightarrow$};
\node[posl, anchor=north] at ([yshift=-2pt]g1.south) {$\alpha_q$};
\end{scope}
\begin{scope}[xshift=27mm]
\node (i2) {$\rightarrow$}
  child { node (p2) {$p$} child { node (pv2) {$v$} } }
  child { node (q2) {$q$} child { node (qv2) {$v$} } };
\begin{pgfonlayer}{background}
\node[fragment=frPeach, fit=(i2)(q2)(qv2)] (f2) {};
\node[fragment=frBlue, fit=(p2)(pv2)] (g2) {};
\end{pgfonlayer}
\node[posl, anchor=west] at (f2.east) {$\gamma_\rightarrow$};
\node[posl, anchor=north] at ([yshift=-2pt]g2.south) {$\beta_p$};
\end{scope}
\end{tikzpicture}

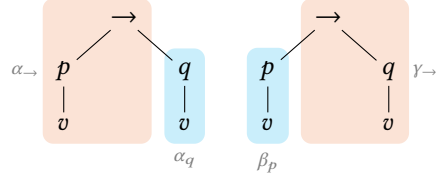
\captionof{figure}{The seed $\psi_\rightarrow$ carved by $\mathcal{L}_1$ (left) and by the extension $\mathcal{L}_2'$ (right). Each colored region is the fragment of the seed covered by one construct. The fragments correspond one to one (same colors), inducing the translation $\alpha_\rightarrow \mapsto \gamma_\rightarrow$, $\alpha_q \mapsto \beta_p$.}
\Description{Two copies of the same formula tree side by side. In the left copy a peach region covers the implication and the antecedent atom, with the consequent atom in a separate blue region; in the right copy a peach region covers the implication and the consequent atom, with the antecedent atom in a separate blue region.}
\label{fig:effective-example}
\end{minipage}
\end{example}

\section{Synthesis of Contrapuntal Languages}
\label{sec:synthesis}

This section solves the \emph{synthesis problem}: given the seed set $\Psi$, find a pair of sugared languages that is effectively contrapuntal for $\Psi$. Rather than reasoning about structure-preserving translations between all pairs of sugarings of every seed, our method enforces a simpler and more conservative requirement: that no two sugarings of a seed have the same \emph{shape}.

\begin{definition}[Shape]
\label{def:shape}
The \emph{shape} of a formula $\chi$ is its set of positions $\mathrm{Pos}(\chi)$; formulas $\chi_1 \in \mathcal{L}_1$ and $\chi_2 \in \mathcal{L}_2$ \emph{have the same shape} iff $\mathrm{Pos}(\chi_1) = \mathrm{Pos}(\chi_2)$---that is, their trees are identical up to labels.
\end{definition}

Formulas related by a structure-preserving translation have the same shape, but not conversely: $p(u) \land p(v)$ and $p(u) \land q(v)$ have the same shape, yet no translation relates them, since it would have to map $p$ to both $p$ and $q$. Forbidding same-shaped sugarings therefore entails clause~(ii) of Definition~\ref{def:effective}, and avoids quantifying over symbol maps: whether two sugarings have the same shape depends only on their positions, which Section~\ref{sec:synth-shape} encodes as quantifier-free constraints.

Our method searches over \emph{cuts} of the seed trees. A cut of a formula is a non-root position $\pi$; it severs the edge above the subtree at $\pi$. A \emph{cut set} $C \subseteq \mathrm{Pos}(\psi) \setminus \{\varepsilon\}$ partitions the tree into \emph{fragments}, one rooted at $\varepsilon$ and one at each cut, each extending downward until it meets a cut or a leaf. It thereby determines a sugar: each fragment becomes the definition of a construct, with its cuts as holes, and the resulting \emph{fragment tree} is a sugared formula that desugars to the seed (Figure~\ref{fig:shape-example}).

\subsection{Incremental Synthesis Algorithm}
\label{sec:synth-algorithm}

\begin{wrapfigure}{R}{0.52\textwidth}
\vspace{-1.1\baselineskip}
\begin{minipage}{\linewidth}
\vspace{-1mm}
\begin{algorithm}[H]
\caption{Incremental synthesis of an effectively contrapuntal pair of sugared languages}
\label{alg:synthesis}
\small
\begin{algorithmic}[1]
\Require a finite, nonempty list $\Psi = \psi_1, \dots, \psi_n$ of seed formalizations in $\mathcal{L}_0$
\Ensure a pair effectively contrapuntal for $\Psi$, or $\mathsf{none}$
\State \Return $\textsc{Synth}(\Psi, \emptyset, \emptyset, \emptyset)$
\Statex
\Function{Synth}{$S, P, \mathcal{L}_1, \mathcal{L}_2$}
  \If{$S$ is empty} \Return $(\mathcal{L}_1, \mathcal{L}_2)$ \EndIf
  \State $\psi \gets$ head of $S$
  \State $B \gets \emptyset$
  \Loop
    \State $\delta \gets \textsc{Cut}(\psi, \mathcal{L}_1, \mathcal{L}_2, B)$ \label{ln:cut}
    \State \textbf{if} $\delta = \mathsf{none}$ \textbf{then return} $\mathsf{none}$
    \State $(\Delta\mathcal{L}_1, \Delta\mathcal{L}_2) \gets \delta$
    \State $(\mathcal{L}_1', \mathcal{L}_2') \gets (\mathcal{L}_1 \cup \Delta\mathcal{L}_1,\, \mathcal{L}_2 \cup \Delta\mathcal{L}_2)$
    \State $w \gets \textsc{FindIso}(P \cup \{\psi\}, \mathcal{L}_1', \mathcal{L}_2')$ \label{ln:verify}
    \If{$w = \mathsf{none}$}
      \State $r \gets \textsc{Synth}(\text{tail of } S,\, P \cup \{\psi\},\, \mathcal{L}_1', \mathcal{L}_2')$ \label{ln:rec}
      \State \textbf{if} $r \neq \mathsf{none}$ \textbf{then return} $r$
    \EndIf
    \State $B \gets B \cup \{\delta\}$ \label{ln:block}
  \EndLoop
\EndFunction
\end{algorithmic}
\end{algorithm}
\end{minipage}
\vspace{-3\baselineskip}
\end{wrapfigure}
The algorithm grows the two languages by adding constructs. An \emph{extension} for a seed $\psi$ over a pair $(\mathcal{L}_1, \mathcal{L}_2)$ is a pair $\delta = (\Delta\mathcal{L}_1, \Delta\mathcal{L}_2)$ of sets of fresh constructs, obtained from two cut sets of $\psi$ whose fragment trees differ in shape: $\Delta\mathcal{L}_i$ defines every fragment of the $i$-th cut set not already in $\mathcal{L}_i$, so that its fragment tree is an $(\mathcal{L}_i \cup \Delta\mathcal{L}_i)$-formula desugaring to $\psi$. Extending thus only adds constructs and never changes existing definitions. Algorithm~\ref{alg:synthesis} is a backtracking fold over the seeds: each invocation of \textsc{Synth} integrates the head of the list $S$ of remaining seeds by one extension and recurses on the tail, keeping its own set $B$ of extensions it has tried and rejected; $P$ is the set of integrated seeds, and both languages start empty.

The algorithm uses two subroutines. $\textsc{Cut}(\psi, \mathcal{L}_1, \mathcal{L}_2, B)$ returns an extension for $\psi$ over $(\mathcal{L}_1, \mathcal{L}_2)$ that is not in $B$, or $\mathsf{none}$ once its search space is exhausted (Section~\ref{sec:synth-cut}). $\textsc{FindIso}(\Psi', \mathcal{L}_1, \mathcal{L}_2)$ returns a witness $(\psi', \chi_1, \chi_2)$, consisting of a seed $\psi' \in \Psi'$ and same-shaped sugarings $\chi_1 \in \Sug_{\mathcal{L}_1}(\psi')$ and $\chi_2 \in \Sug_{\mathcal{L}_2}(\psi')$, or $\mathsf{none}$ if there is none (Section~\ref{sec:synth-findiso}). This check is needed because an extension guarantees different shapes only for the two fragment trees it was built from: constructs may recombine into other sugarings of a seed, as $\gamma_\rightarrow$ recombines with $\beta_p$ in Example~\ref{ex:effective}. New constructs can create such recombinations for earlier seeds too, so line~\ref{ln:verify} rechecks every integrated seed.

An extension is blocked (line~\ref{ln:block}) either because it violates the shape constraint, when \textsc{FindIso} returns a witness, or because it is a dead end, when the recursive call finds no completion over the remaining seeds; the loop then asks \textsc{Cut} for another extension. Since $B$ is local to the invocation, it starts empty whenever the languages are extended and is restored on backtracking.

\begin{theorem}[Correctness]
\label{thm:correctness}
Algorithm~\ref{alg:synthesis} terminates, and if it returns a pair of languages, the pair is effectively contrapuntal for $\Psi$.
\end{theorem}

\noindent The proof is given in our supplementary materials. The algorithm is incomplete: it may return $\mathsf{none}$ even if an effectively contrapuntal pair exists.

\subsection{Encoding Shape Constraints}
\label{sec:synth-shape}

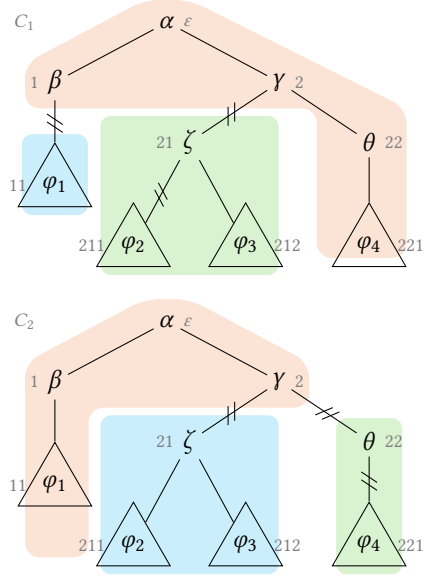
\begin{wrapfigure}{r}{0.4\textwidth}
\vspace{-0.3\baselineskip}
\centering
\resizebox{\linewidth}{!}{%
\begin{tikzpicture}[
  level distance=8mm,
  level 1/.style={sibling distance=30mm},
  level 2/.style={sibling distance=24mm},
  level 3/.style={sibling distance=15mm},
  subtree/.append style={minimum size=6mm, anchor=north},
  every node/.style={inner sep=2pt}
]
\begin{scope}
\node[posl] at (-19mm,0mm) {$C_1$};
\node (r1) [label={[posl]right:$\varepsilon$}] {$\alpha$}
  child { node (b1) [label={[posl]left:$1$}] {$\beta$}
    child { node[subtree] (f11) [label={[posl]left:$11$}] {$\varphi_1$} edge from parent[cut] } }
  child { node (g1) [label={[posl]right:$2$}] {$\gamma$}
    child { node (z1) [label={[posl]left:$21$}] {$\zeta$}
      child { node[subtree] (f21) [label={[posl]left:$211$}] {$\varphi_2$} edge from parent[cut] }
      child { node[subtree] (f31) [label={[posl]right:$212$}] {$\varphi_3$} }
      edge from parent[cut] }
    child { node (e1) [label={[posl]right:$22$}] {$\theta$}
      child { node[subtree] (f41) [label={[posl]right:$221$}] {$\varphi_4$} } } };
\begin{pgfonlayer}{background}
\node[fragment=frBlue, fit=(f11)] {};
\node[fragment=frGreen, fit=(z1)(f21)(f31)] {};
\fill[frPeach, rounded corners=2mm]
  (-19mm,-11.5mm) -- (-19mm,-5.5mm) -- (-3mm,3mm) -- (3mm,3mm) -- (32mm,-11.5mm) -- (32mm,-31.5mm) -- (20mm,-31.5mm) -- (20mm,-11.5mm) -- cycle;
\end{pgfonlayer}
\end{scope}
\begin{scope}[yshift=-40mm]
\node[posl] at (-19mm,0mm) {$C_2$};
\node (r2) [label={[posl]right:$\varepsilon$}] {$\alpha$}
  child { node (b2) [label={[posl]left:$1$}] {$\beta$}
    child { node[subtree] (f12) [label={[posl]left:$11$}] {$\varphi_1$} } }
  child { node (g2) [label={[posl]right:$2$}] {$\gamma$}
    child { node (z2) [label={[posl]left:$21$}] {$\zeta$}
      child { node[subtree] (f22) [label={[posl]left:$211$}] {$\varphi_2$} }
      child { node[subtree] (f32) [label={[posl]right:$212$}] {$\varphi_3$} }
      edge from parent[cut] }
    child { node (e2) [label={[posl]right:$22$}] {$\theta$}
      child { node[subtree] (f42) [label={[posl]right:$221$}] {$\varphi_4$} edge from parent[cut] }
      edge from parent[cut] } };
\begin{pgfonlayer}{background}
\node[fragment=frBlue, fit=(z2)(f22)(f32)] {};
\node[fragment=frGreen, fit=(e2)(f42)] {};
\fill[frPeach, rounded corners=2mm]
  (-19mm,-31.5mm) -- (-19mm,-5.5mm) -- (-3mm,3mm) -- (3mm,3mm) -- (19mm,-5.5mm) -- (19mm,-11.5mm) -- (-10.5mm,-11.5mm) -- (-10.5mm,-31.5mm) -- cycle;
\end{pgfonlayer}
\end{scope}
\end{tikzpicture}}
\caption{Two cut sets over the same seed: $C_1 = \{11, 21, 211\}$ and $C_2 = \{21, 22, 221\}$. Gray labels give positions, triangles stand for subtrees, and double strokes mark cuts; colors mark the root fragment (peach) and the subtrees below it (blue, green). Both fragment trees have the shape $\{\varepsilon, 1, 2, 21\}$.}
\Description{Two copies of the same schematic tree, one above the other. In the upper copy, cut marks cross the edge into the left subtree triangle and, in the right branch, the edge into an inner node and one edge below it. In the lower copy, cut marks cross, in the right branch only, the edges into two inner nodes and one edge below the second of them.}
\label{fig:shape-example}
\vspace{-6\baselineskip}
\end{wrapfigure}

Figure~\ref{fig:shape-example} shows two cut sets that cut the same seed in unrelated places, yet yield fragment trees of the same shape. What matters is not where the cuts lie but how they nest: the holes of a fragment are its \emph{dominating} cuts, those below its root with no other cut in between. At the root, $C_1$ has the dominating cuts $11$ and $21$ (with $211$ below $21$), and $C_2$ has $21$ and $22$ (with $221$ below $22$), so both top constructs have two holes. The constraints below capture this condition node by node.

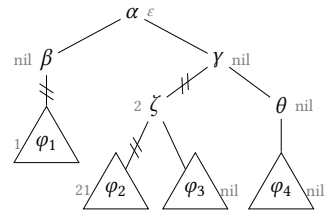
\begin{wrapfigure}{r}{0.3\textwidth}
\vspace{-\baselineskip}
\centering
\resizebox{\linewidth}{!}{%
\begin{tikzpicture}[
  level distance=7mm,
  level 1/.style={sibling distance=26mm},
  level 2/.style={sibling distance=19mm},
  level 3/.style={sibling distance=12mm},
  subtree/.append style={minimum size=6mm, anchor=north},
  every node/.style={inner sep=2pt}
]
\node (ar) [label={[posl]right:$\varepsilon$}] {$\alpha$}
  child { node (ab) [label={[posl]left:$\mathrm{nil}$}] {$\beta$}
    child { node[subtree] (af1) [label={[posl]left:$1$}] {$\varphi_1$} edge from parent[cut] } }
  child { node (ag) [label={[posl]right:$\mathrm{nil}$}] {$\gamma$}
    child { node (az) [label={[posl]left:$2$}] {$\zeta$}
      child { node[subtree] (af2) [label={[posl]left:$21$}] {$\varphi_2$} edge from parent[cut] }
      child { node[subtree] (af3) [label={[posl]right:$\mathrm{nil}$}] {$\varphi_3$} }
      edge from parent[cut] }
    child { node (ae) [label={[posl]right:$\mathrm{nil}$}] {$\theta$}
      child { node[subtree] (af4) [label={[posl]right:$\mathrm{nil}$}] {$\varphi_4$} } } };
\end{tikzpicture}}
\caption{The array assignment for $C_1$ (Figure~\ref{fig:shape-example}); gray annotations give $a^1_\pi$.}
\Description{The upper tree of the previous figure, with the same three cut marks, where every node is annotated by an array value: the root by the empty position, the three nodes below cuts by the positions 1, 2, and 21, and all remaining nodes by nil.}
\label{fig:shape-assignment}
\vspace{-\baselineskip}
\end{wrapfigure}

We encode a pair of cut sets $C_1, C_2$ of a seed $\psi$ symbolically, one for each language. Let $i \in \{1, 2\}$, and let all positions below range over $\mathrm{Pos}(\psi)$. For every non-root position $\pi$, a Boolean \emph{cut variable} $c^i_\pi$ indicates that $\pi \in C_i$. For every position $\pi$, a symbolic array $a^i_\pi$ records where the node at $\pi$ ends up in the fragment tree of $C_i$: if $\pi$ roots a construct ($\pi = \varepsilon$ or $\pi \in C_i$), $a^i_\pi$ holds its position, and otherwise the constant $\mathrm{nil}$; we also write $\varepsilon$ for the empty array. A construct's position gains one element per cut on the path from the root, so $a^i_\pi$ has no more elements than $\pi$. Figure~\ref{fig:shape-assignment} shows the arrays for $C_1$. The non-$\mathrm{nil}$ arrays of a cut set thus enumerate the positions of its fragment tree, which is how the encoding compares shapes.

Three families of constraints determine the arrays from the cuts, a fourth compares the two cuttings, and a fifth handles binders. We write $a^i_\pi \sqsubset_1 a^i_\rho$ if neither array is $\mathrm{nil}$ and $a^i_\rho$ holds a child of the position held by $a^i_\pi$.

The \emph{definedness constraints} place the root of the seed at position $\varepsilon$ of the fragment tree and make the array of any other position non-$\mathrm{nil}$ exactly when that position is cut:
\[
\mathrm{Def}^i \;\triangleq\; a^i_\varepsilon \doteq \varepsilon \;\land \bigwedge_{\pi \neq \varepsilon} \bigl(\, a^i_\pi \not\doteq \mathrm{nil} \,\leftrightarrow\, c^i_\pi \bigr).
\]

The \emph{domination constraints} determine the edges of the fragment tree: the construct at $\rho$ is a child of the construct at $\pi$ exactly when $\rho$ is cut and no cut lies strictly between $\pi$ and $\rho$:
\[
\mathrm{Dom}^i \;\triangleq\; \bigwedge_{\pi \sqsubset \rho} \Bigl(\, a^i_\pi \sqsubset_1 a^i_\rho \;\leftrightarrow\; a^i_\pi \not\doteq \mathrm{nil} \,\land\, c^i_\rho \,\land \bigwedge_{\pi \sqsubset \sigma \sqsubset \rho} \lnot c^i_\sigma \Bigr).\hspace{0.45\textwidth}
\]

Domination determines which constructs are children of which, but not in what order. The \emph{ordering constraints} number the children of every construct $1, \dots, d$ from left to right, following the order of their cuts in the seed. We write $\rho \prec \rho'$ if $\rho$ lies to the left of $\rho'$ (neither is a prefix of the other, and $\rho$ is smaller at the first index where they differ); $\mathrm{sib}(a^i_\rho, a^i_{\rho'})$ if both arrays are non-$\mathrm{nil}$ and hold positions of equal length that differ at most in the last element; and $\mathrm{last}(a^i_\rho)$ for the last element, where comparisons on $\mathrm{last}$ imply that the arrays are non-$\mathrm{nil}$ and nonempty:
\par\noindent\begin{minipage}{0.56\textwidth}
\begin{align*}
\mathrm{Ord}^i \;\triangleq\; &\bigwedge_{\rho \prec \rho'} \Bigl(\, \mathrm{sib}(a^i_\rho, a^i_{\rho'}) \;\rightarrow\; \mathrm{last}(a^i_\rho) < \mathrm{last}(a^i_{\rho'}) \Bigr)\\
{}\land\; &\bigwedge_{\rho} \Bigl(\, \mathrm{last}(a^i_\rho) > 1 \;\rightarrow\\
&\quad \bigvee_{\rho' \prec \rho} \bigl(\, \mathrm{sib}(a^i_{\rho'}, a^i_\rho) \,\land\, \mathrm{last}(a^i_\rho) \doteq \mathrm{last}(a^i_{\rho'}) + 1 \bigr) \Bigr).
\end{align*}
\end{minipage}\par\noindent
The first conjunct orders siblings as their cuts are ordered in the seed and, since $\mathrm{sib}$ allows equal last elements, prevents two constructs from sharing a position; the second makes indices contiguous.

The \emph{same-shape constraint} requires every construct position realized in one fragment tree to be realized in the other:
\[
\mathrm{Shape} \;\triangleq\; \bigwedge_{\pi} \Bigl(\, a^1_\pi \not\doteq \mathrm{nil} \;\rightarrow\; \bigvee_{\rho}\, a^1_\pi \doteq a^2_\rho \Bigr) \;\land\; \bigwedge_{\rho} \Bigl(\, a^2_\rho \not\doteq \mathrm{nil} \;\rightarrow\; \bigvee_{\pi}\, a^2_\rho \doteq a^1_\pi \Bigr).
\]
Since $\mathrm{Def}^i \land \mathrm{Dom}^i \land \mathrm{Ord}^i$ force the arrays to hold the actual positions of the constructs, $\mathrm{Shape}$ holds exactly when the two fragment trees have the same shape.

The \emph{binding constraints} keep every bound variable occurrence in the same fragment as its quantifier, as sugarings require (Section~\ref{sec:bg-sugar}); for example, they forbid cutting $p(v)$ out of $\forall v.\,(p(v) \rightarrow q(v))$. We write $\pi \mathrel{\mathrm{binds}} \rho$ if $\pi$ is labeled $\forall x$ or $\exists x$ and $\rho$ is an occurrence of $x$ bound by it:
\[
\mathrm{Bnd}^i \;\triangleq\; \bigwedge_{\pi \,\mathrm{binds}\, \rho}\; \bigwedge_{\pi \sqsubset \sigma \sqsubseteq \rho} \lnot c^i_\sigma.\hspace{0.32\textwidth}
\]

\subsection{Subroutines \textsc{Cut} and \textsc{FindIso}}
\label{sec:synth-cut}

Both subroutines search cut sets of a seed with the constraints of Section~\ref{sec:synth-shape}: \textsc{Cut} for two cut sets whose fragment trees differ in shape, and \textsc{FindIso} for two sugarings of the same shape built from existing constructs. We solve all queries, including a \textsc{MaxSat} problem, with the Z3 SMT solver~\cite{demoura2008z3}.

\begin{wrapfigure}{r}{0.44\textwidth}
\vspace{-0.3\baselineskip}
\centering
\resizebox{\linewidth}{!}{%
\begin{tikzpicture}[
  level distance=8mm,
  every node/.style={inner sep=2pt},
  subtree/.append style={minimum size=6mm, anchor=north}
]
\begin{scope}[level 1/.style={sibling distance=13mm}]
\node[posl, anchor=east] at (-6mm, 0mm) {$D_{\kappa_1}$};
\node (k1) {$\beta$}
  child { node (kz) {$\zeta$} child { node {$\square_1$} } }
  child { node {$\square_2$} };
\begin{pgfonlayer}{background}
\node[fragment=frBlue, fit=(k1)(kz)] {};
\end{pgfonlayer}
\end{scope}
\begin{scope}[yshift=-24mm]
\node[posl, anchor=east] at (-6mm, 0mm) {$D_{\kappa_2}$};
\node (k2) {$\theta$}
  child { node {$\square_1$} };
\begin{pgfonlayer}{background}
\node[fragment=frGreen, fit=(k2)] {};
\end{pgfonlayer}
\end{scope}
\begin{scope}[xshift=38mm, level 1/.style={sibling distance=22mm}, level 2/.style={sibling distance=13mm}]
\node (sr) [label={[posl]right:$\varepsilon$}] {$\alpha$}
  child { node (sb) {$\beta$}
    child { node (sz) {$\zeta$}
      child { node[subtree] (sf1) {$\varphi_1$} edge from parent[cut] } }
    child { node[subtree] (sf2) {$\varphi_2$} edge from parent[cut] }
    edge from parent[cut] }
  child { node (sg) {$\gamma$}
    child { node (se) {$\theta$}
      child { node[subtree] (sf3) {$\varphi_3$} edge from parent[cut] }
      edge from parent[cut] }
    child { node[subtree] (sf4) {$\varphi_4$} } };
\begin{pgfonlayer}{background}
\node[fragment=frBlue, fit=(sb)(sz)] (m1) {};
\node[fragment=frGreen, fit=(se)] (m2) {};
\end{pgfonlayer}
\node[posl, anchor=east] at (m1.west) {$\kappa_1$};
\node[posl, anchor=east] at (m2.west) {$\kappa_2$};
\end{scope}
\end{tikzpicture}}
\caption{Matching constructs $\kappa_1, \kappa_2 \in \Lambda_1$ (definitions on the left, colored as their occurrences) against a seed (right). Each adopted match cuts at the root and holes of its occurrence but not inside it; the uncovered remainder ($\alpha$, $\gamma$, $\varphi_4$) is left to the cutting step.}
\Description{Two small trees with square hole leaves on the left, and a seed tree on the right. In the seed, two colored regions, shaded like the two small trees, mark the occurrences of their definitions; cut marks cross the edges entering each contour and the edges leaving them toward subtree triangles, while the root, one inner node, and one subtree triangle lie outside both contours.}
\label{fig:matching}
\vspace{-1.7\baselineskip}
\end{wrapfigure}
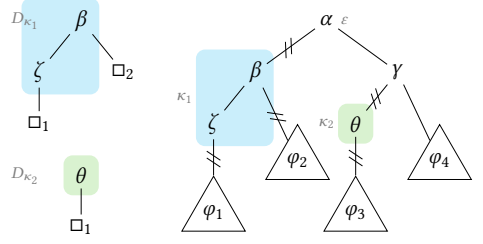
$\textsc{Cut}(\psi, \mathcal{L}_1, \mathcal{L}_2, B)$ works in two steps: a \emph{matching} step decides which constructs of $\mathcal{L}_1$ and $\mathcal{L}_2$ are reused in $\psi$, and a \emph{cutting} step cuts the rest of $\psi$.

A \emph{match} of a construct $\alpha \in \Lambda_i$ at a position $\pi$ is an occurrence of $D_\alpha$ in $\psi$: its non-hole nodes coincide with the nodes of $\psi$ at and below $\pi$, and its holes fall at positions $\rho_1, \dots, \rho_{\mathit{ar}(\alpha)}$. Adopting the match makes the occurrence one fragment:
\[
\mathrm{match}^i_{\alpha, \pi} \;\triangleq\; c^i_\pi \;\land \bigwedge_{j} c^i_{\rho_j} \;\land \bigwedge_{\sigma} \lnot c^i_\sigma,\hspace{0.45\textwidth}
\]
where $\sigma$ ranges over the non-hole positions strictly below $\pi$ covered by $D_\alpha$, and $c^i_\pi$ is dropped when $\pi = \varepsilon$. Matches are found by syntactic comparison, at most one per construct and position.

The matching step reuses as many constructs of $\mathcal{L}_1$ and $\mathcal{L}_2$ as possible. Adopting every match is impossible in general, since two matches may disagree on a cut variable, so we treat the match formulas as soft constraints and use \textsc{MaxSat} to find a largest set of them that hold together with $\mathrm{Bnd}$. The cut variables they determine are then fixed (Figure~\ref{fig:matching}).

The cutting step then solves
\[
\mathrm{Def} \land \mathrm{Dom} \land \mathrm{Ord} \land \lnot\mathrm{Shape} \land \mathrm{Bnd} \land \mathrm{Block}
\]
where $\mathrm{Def} \triangleq \mathrm{Def}^1 \land \mathrm{Def}^2$, and likewise for the other families. The cut variables fixed by matching are frozen, so the solver chooses cuts only in the unmatched remainder of $\psi$, which keeps the search of Algorithm~\ref{alg:synthesis} tractable. Within this space, $\mathrm{Def}$, $\mathrm{Dom}$, $\mathrm{Ord}$, and $\mathrm{Bnd}$ describe the two cuttings, $\lnot\mathrm{Shape}$ requires their fragment trees to differ in shape, and $\mathrm{Block}$ rules out the extensions already tried: since $\psi$, $\mathcal{L}_1$, and $\mathcal{L}_2$ are fixed across the calls of one loop, an extension is determined by its pair of cut sets, so, writing $(C_1^\delta, C_2^\delta)$ for the cut sets of an extension $\delta \in B$,
\[
\mathrm{Block} \;\triangleq\; \bigwedge_{\delta \in B}\, \lnot \bigwedge_{i \in \{1,2\}} \Bigl(\, \bigwedge_{\pi \in C_i^\delta} c^i_\pi \;\land \bigwedge_{\pi \notin C_i^\delta} \lnot c^i_\pi \Bigr).
\]
A satisfying assignment of the query yields the extension directly: its cut variables give the two cut sets, fragments that are definitions of existing constructs reuse them, and every other fragment becomes a fresh construct of $\Delta\mathcal{L}_1$ or $\Delta\mathcal{L}_2$. If the query is unsatisfiable, \textsc{Cut} returns $\mathsf{none}$.

\label{sec:synth-findiso}$\textsc{FindIso}(\Psi', \mathcal{L}_1, \mathcal{L}_2)$ uses the same ingredients in the opposite direction: for each seed $\psi' \in \Psi'$, it computes the matches of the constructs of $\mathcal{L}_1$ and $\mathcal{L}_2$ in $\psi'$ and asks whether some pair of sugarings has the same shape.

A sugaring of $\psi'$ in $\mathcal{L}_i$ is a \emph{tiling} of $\psi'$ by occurrences of constructs of $\mathcal{L}_i$. We introduce an \emph{adoption} variable $m^i_{\alpha, \pi}$ for every match and require the adopted matches to commit their cut variables and, together, to cover $\psi'$:
\[
\mathrm{Tile}^i \;\triangleq\; \bigwedge_{\alpha, \pi} \bigl(\, m^i_{\alpha, \pi} \rightarrow \mathrm{match}^i_{\alpha, \pi} \bigr) \;\land\; \bigwedge_{\sigma}\, \bigvee_{(\alpha, \pi) \ni \sigma} m^i_{\alpha, \pi},
\]
where $(\alpha, \pi) \ni \sigma$ means that $\sigma$ is a non-hole position of the occurrence of $D_\alpha$ at $\pi$. Coverage ensures that every cut stems from a match: since match formulas keep interiors uncut, a cut position can be covered only as the root of an adopted occurrence.

For each $\psi' \in \Psi'$, \textsc{FindIso} then checks
\[
\mathrm{Tile}^1 \land \mathrm{Tile}^2 \land \mathrm{Def} \land \mathrm{Dom} \land \mathrm{Ord} \land \mathrm{Shape} \land \mathrm{Bnd},
\]
now asserting $\mathrm{Shape}$ positively. The matches adopted in a satisfying assignment form same-shaped sugarings $\chi_1 \in \Sug_{\mathcal{L}_1}(\psi')$ and $\chi_2 \in \Sug_{\mathcal{L}_2}(\psi')$, giving the witness $(\psi', \chi_1, \chi_2)$; if the query is unsatisfiable for every $\psi' \in \Psi'$, \textsc{FindIso} returns $\mathsf{none}$.

\section{Evaluation}
\label{sec:evaluation}

Our experiments address the following research questions:
\begin{description}
\item[RQ1:] How effective are contrapuntal languages at validating autoformalized specifications compared with baselines?
\item[RQ2:] (Ablation) How does contrapuntality affect the validation performance?
\item[RQ3:] Do contrapuntal languages help mitigate correlated errors?
\item[RQ4:] What is the token cost of contrapuntal languages in comparison with baselines?
\item[RQ5:] What are the causes of failures of contrapuntal languages?
\end{description}

\paragraph*{\textbf{Datasets.}}
We evaluate on two programming benchmarks. The first is \textbf{\emph{HumanEval+}} (HEP)~\cite{liu2023your}, whose tasks provide a natural-language description, a Python signature, a reference implementation, and tests. Although LLMs solve most HumanEval+ tasks, prior studies show that generating behaviorally correct specifications for HumanEval remains challenging even when expressed as Python postconditions and contracts~\cite{endres2024largelanguagemodelstransform,richter2025postconditionslargelanguagemodels}. Our results show that this challenge persists when formalizing these tasks in SMT-LIB: even with reasoning enabled, Qwen3.6-Flash achieves an average pass@30 below 50\%. For testing, we use HumanEval+-mini~\cite{liu2023your}, which EvalPlus derives from the full suite by a greedy set cover that preserves branch coverage, mutant killing, and the detection of incorrect LLM-generated programs. It is about $47\times$ smaller with similar testing effectiveness, which substantially reduces the cost of SMT queries.

The second is \textbf{\emph{LiveCodeBench}} (LCB)~\cite{jain2025livecodebench}, whose competitive programming problems from AtCoder and LeetCode are substantially harder to formalize. Applying the same test-suite reduction, we construct LiveCodeBench-mini: 727 of the original 7,000 tests across 175 tasks (463 public and 264 private), which still exercise all 6,340 branch arcs covered by the validated reference implementations and detect all 2,587 functionally exposed mutants considered during test selection.

\paragraph*{\textbf{Models.}}
We evaluate using three LLM settings. \textbf{\emph{GPT-5.4 mini}} is a compact proprietary OpenAI model with strong coding, reasoning, and tool-use capabilities. \textbf{\emph{Qwen3.6-Flash}} is a fast-serving model in Qwen family, which includes open-weight releases and emphasizes multilingual coding. This model supports both \textbf{thinking mode enabled} and \textbf{thinking mode disabled}. We conducted experiments under these two modes, respectively.

\paragraph*{\textbf{Methods.}}
For each task, we prompt the LLM with the task description, function signature, and types to generate an SMT-LIB v2 specification. It declares constants for the arguments and return values and contains two named assertions: \texttt{pre\_condition}, for valid inputs, and \texttt{post\_condition}, for the input--output relation. Since we focus on validating autoformalized specifications: deciding which candidates to accept when no external correctness oracle is available, we therefore compare three such strategies using the same sampling budget of 30 LLM-generated responses per task, and include direct generation as a reference to assess the effect of validation.

\textbf{\emph{Direct generation}} keeps all $n=30$ samples that pass basic validity checks, i.e., follow the pre-/postcondition structure and parse in Z3 (We discuss the choice of $n$ in supplementary materials). 

\textbf{\emph{Plurality voting}} follows the Symbolic Equivalence strategy~\cite{NEURIPS2024_6034a661}, clusters these specifications by pairwise SMT equivalence and selects the largest class. 

\textbf{\emph{Round-trip veri.}} follows the round-trip verification strategy~\cite{amrollahi2026faithfulautoformalizationroundtripverification}, translating each of $n = 10$ valid seeds into natural language and back into SMT-LIB. It retains only seeds proven SMT-equivalent to their regenerated counterparts, which serve solely as verification artifacts. 

\textbf{\emph{Contrapuntal langs.}} synthesizes a pair of contrapuntal languages from $n = 10$ valid seeds and generates $m = 10$ specifications in each language; a candidate is kept if it passes the validity checks, uses the constructs of its language, and is proven equivalent to a candidate from the other language.

\paragraph*{\textbf{Mutant generation.}}
To assess soundness, we need incorrect programs, called mutants, whose outputs a sound specification must reject. Since HumanEval+ provides no incorrect programs, we prompt GPT-5.5 with the canonical solution to produce distinct single-bug variants, targeting five per task; for LiveCodeBench, we combine real wrong-answer submissions with at least five GPT-5.5 variants of a validated correct submission. We discard candidates that are syntactically invalid, lack the required entry point, or duplicate others, and run the rest on the benchmark tests. A mutant yields a negative example only when it returns a value different from the reference output; exceptions and resource failures yield no input--output pair to check against a specification and are ignored. This gives 779 mutants for HumanEval+ (161 tasks, 4.84 per task) and 2,587 for LiveCodeBench (175 tasks, 14.78 per task: 1,618 real submissions and 969 generated variants).

\paragraph*{\textbf{Evaluation Method.}}
Neither benchmark provides reference formal specifications, so, following prior behavioral evaluations~\cite{endres2024largelanguagemodelstransform,richter2025postconditionslargelanguagemodels,prasetya2026talk}, we evaluate specifications on input--output pairs: \emph{positive tests} from reference executions and \emph{negative tests} from mutant outputs, all with inputs valid for their tasks. For each pair, Z3 checks whether the specification's pre- and postconditions, with the variables bound to the pair, are satisfiable, that is, whether the specification admits the pair. A positive test passes if its pair is admitted; rejecting it, including through the precondition, is a completeness failure. A negative test kills its mutant if the pair is rejected while the input satisfies the precondition, so that an overly restrictive precondition does not count as a kill. A mutant is killed if any of its negative tests kills it, and a specification is \emph{sound} if it kills every mutant.

Since the solver may time out, we use a threshold $\tau \in [0, 1]$: a specification is \emph{$\tau$-complete} if no positive test fails, at least a fraction $\tau$ pass, and the rest return \textsc{unknown}. An \textsc{unknown} result counts as neither a pass nor a kill; errored or unsupported cases are excluded.

\subsection{RQ1: Effectiveness of Specification Validation}
\label{sec:rq1}

Reporting either completeness or soundness alone for specifications accepted under different methods can be misleading, since a specification may trivially favor completeness by accepting all behaviors, or favor soundness by rejecting all behaviors. We therefore report two joint metrics: MKR$\mid$C denotes the percentage of evaluated mutants killed by accepted specifications that are $\tau$-complete; CR denotes the empirical correctness rate: the percentage of accepted specifications that are both $\tau$-complete and sound (able to kill all evaluated mutants). Their task-level macro-averages are denoted by Avg.~MKR$\mid$C and Avg.~CR, respectively. The discussion below focuses on $\tau=0.8$ and $\tau=1.0$; In our supplementary materials, we present a sensitivity analysis across the full range $\tau\in[0,1]$.

\begin{table}[t]
\captionsetup{aboveskip=3pt, belowskip=3pt}
\caption{Validation results for accepted specifications. Higher is better; bold marks the best result for each metric and threshold within each benchmark (BM) and model configuration.}
\label{tab:hep_result}
\centering
\footnotesize
\begin{threeparttable}
\renewcommand{\arraystretch}{0.85}
\begin{tabularx}{\linewidth}{lll *{8}{>{\centering\arraybackslash}X}}
\toprule
\multirow{2}{*}{\textbf{BM}}
& \multirow{2}{*}{\textbf{Model}}
& \multirow{2}{*}{\textbf{Method}}
& \multicolumn{2}{c}{\makebox[0pt][c]{\textbf{MKR$\mid$C (\%)}}}
& \multicolumn{2}{c}{\makebox[0pt][c]{\textbf{CR (\%)}}}
& \multicolumn{2}{c}{\makebox[0pt][c]{\textbf{Avg. MKR$\mid$C (\%)}}}
& \multicolumn{2}{c}{\makebox[0pt][c]{\textbf{Avg. CR (\%)}}} \\
& &
& \textbf{$\tau$=0.8} & \textbf{$\tau$=1.0}
& \textbf{$\tau$=0.8} & \textbf{$\tau$=1.0}
& \textbf{$\tau$=0.8} & \textbf{$\tau$=1.0}
& \textbf{$\tau$=0.8} & \textbf{$\tau$=1.0} \\
\midrule
\multirow{12}{*}{\textbf{HEP}}
& \multirow{4}{*}{\textbf{GPT-5.4 mini}}
& Direct generation
& 79.52 & 77.75 & 21.41 & 18.30 & 59.17 & 55.62 & 19.66 & 16.70 \\
& & Round-trip Veri.
& 84.35 & 82.67 & 28.30 & 23.78 & 73.80 & 71.24 & 21.62 & 18.72 \\
& & Plurality voting
& 92.26 & 91.25 & 41.69 & 35.59 & 81.22 & 80.44 & 23.81 & 21.24 \\
& & Contrapuntal langs.
& \textbf{99.04} & \textbf{98.86} & \textbf{52.30} & \textbf{43.54}
& \textbf{98.89} & \textbf{98.60} & \textbf{47.52} & \textbf{37.59} \\
\addlinespace
& \multirow{4}{*}{\shortstack{\textbf{Qwen3.6-Flash} \\ thinking=off}}
& Direct generation
& 85.95 & 85.93 & 37.91 & 34.94 & 71.35 & 69.78 & 24.43 & 22.67 \\
& & Round-trip Veri.
& 88.39 & 89.72 & 59.78 & 53.63 & 84.64 & 85.18 & 43.06 & 39.62 \\
& & Plurality voting
& 91.55 & 92.52 & 62.70 & 58.81 & 83.12 & 84.67 & 37.31 & 35.07 \\
& & Contrapuntal langs.
& \textbf{98.04} & \textbf{98.04} & \textbf{68.47} & \textbf{68.47}
& \textbf{90.67} & \textbf{90.67} & \textbf{50.00} & \textbf{50.00} \\
\addlinespace
& \multirow{4}{*}{\shortstack{\textbf{Qwen3.6-Flash} \\ thinking=on}}
& Direct generation
& 87.76 & 86.65 & 48.99 & 41.76 & 71.82 & 64.39 & 37.37 & 30.85 \\
& & Round-trip Veri.
& 94.64 & 94.27 & 76.03 & \textbf{70.62} & 84.86 & 83.87 & 59.48 & 54.06 \\
& & Plurality voting
& 94.98 & 94.58 & 69.85 & 62.67 & 87.35 & 85.43 & 51.14 & 42.94 \\
& & Contrapuntal langs.
& \textbf{97.86} & \textbf{99.01} & \textbf{81.25} & 68.75
& \textbf{96.06} & \textbf{96.95} & \textbf{76.18} & \textbf{64.91} \\
\midrule
\multirow{12}{*}{\textbf{LCB}}
& \multirow{4}{*}{\textbf{GPT-5.4 mini}}
& Direct generation
& 43.49 & 42.82 & 9.03 & 8.55 & 29.91 & 26.72 & 7.56 & 6.98 \\
& & Round-trip Veri.
& 40.01 & 42.33 & 11.11 & 10.84 & 28.56 & 29.22 & 9.16 & 8.84 \\
& & Plurality voting
& 56.82 & 55.58 & 27.65 & 26.36 & 33.80 & 31.39 & 11.02 & 9.84 \\
& & Contrapuntal langs.
& \textbf{88.70} & \textbf{88.70} & \textbf{50.43} & \textbf{50.43}
& \textbf{76.81} & \textbf{76.81} & \textbf{38.10} & \textbf{38.10} \\
\addlinespace
& \multirow{4}{*}{\shortstack{\textbf{Qwen3.6-Flash} \\ thinking=off}}
& Direct generation
& 63.70 & 68.70 & 21.41 & 19.54 & 35.70 & 37.19 & 9.92 & 8.64 \\
& & Round-trip Veri.
& 79.13 & 88.11 & 33.33 & 32.56 & 57.11 & 65.42 & 20.37 & 17.99 \\
& & Plurality voting
& 78.37 & 83.92 & 44.49 & 43.70 & 50.97 & 60.00 & 23.81 & 21.43 \\
& & Contrapuntal langs.
& \textbf{100.0} & \textbf{100.0} & \textbf{55.56} & \textbf{55.56}
& \textbf{100.0} & \textbf{100.0} & \textbf{42.86} & \textbf{42.86} \\
\addlinespace
& \multirow{4}{*}{\shortstack{\textbf{Qwen3.6-Flash} \\ thinking=on}}
& Direct generation
& 78.51 & 80.93 & 32.06 & 28.12 & 48.50 & 47.91 & 15.03 & 12.47 \\
& & Round-trip Veri.
& 72.75 & 75.07 & 42.68 & 41.06 & 54.41 & 58.68 & 30.65 & 28.61 \\
& & Plurality voting
& 88.59 & 91.23 & 60.83 & 57.91 & 64.92 & 70.88 & 33.33 & 29.24 \\
& & Contrapuntal langs.
& \textbf{100.0} & \textbf{100.0} & \textbf{68.67} & \textbf{63.86}
& \textbf{100.0} & \textbf{100.0} & \textbf{62.50} & \textbf{58.93} \\
\bottomrule
\end{tabularx}
\begin{tablenotes}[flushleft]
\footnotesize

\item[a] $\tau$: the minimum fraction of checked positive tests that must return \textsc{pass} for a complete specification, provided that none of them returns \textsc{fail}. Any remaining checked tests may only return \textsc{unknown} (e.g., due to timeout).

\item[b] \textbf{MKR$\mid$C}: \textbf{M}utant-\textbf{K}ill \textbf{R}ate conditioned on \textbf{C}ompleteness---the percentage of evaluated mutants killed by accepted specifications that are $\tau$-complete.

\item[c] \textbf{CR}: \textbf{C}orrectness \textbf{R}ate---the percentage of accepted specifications that are $\tau$-complete and sound.

\item[d] MKR$\mid$C and CR use sample-level aggregation, weighting all accepted specifications equally. \textbf{Avg.} columns report task-level macro-averages: the corresponding metric is computed within each task and then averaged across tasks with equal weight.
\end{tablenotes}
\end{threeparttable}
\vspace{-3mm}
\end{table}

As shown in Table~\ref{tab:hep_result}, contrapuntal languages achieve significantly higher Avg.~CR than both round-trip verification and plurality voting on average under a paired task-level bootstrap (100,000 resamples). The gains over round-trip verification are 22.13 percentage points at $\tau=0.8$ (approximate 95\% simultaneous confidence interval: $[6.67,37.60]$) and 20.76 points at $\tau=1.0$ ($[5.27,36.24]$); the corresponding gains over plurality voting are 22.79 points ($[7.57,38.00]$) and 22.10 points ($[6.87,37.34]$), respectively. The other metrics in Table~\ref{tab:hep_result} also show improvements: averaged across the three model configurations, the relative gains over plurality voting in MKR$\mid$C, Avg.~MKR$\mid$C, and CR are 5.8--6.3\%, 13.6--14.4\%, and 16.2--17.0\% on HEP, and 29.5--32.2\%, 84.1--92.5\%, and 40.1--42.9\% on LCB, respectively. Contrapuntal languages also outperform round-trip verification in these metrics across all reported settings except CR $(\tau=1.0)$ for Qwen3.6-Flash with thinking on HEP.

The results above cover only tasks with an accepted specification. We therefore also report \emph{task coverage}, the fraction of tasks with at least one accepted specification. For reference, we also report \emph{seed availability}, the fraction of tasks whose initial seed pool contains at least one complete and sound specification. This is an upper bound on the fraction of tasks any method can accept correctly. Seed availability is 39.17\% on HEP and 12.97\% on LCB. Our method covers 29.81\% and 8.76\%, round-trip validation 64.60\% and 52.00\%, and plurality voting 73.27\% and 48.77\%. Both baselines exceed seed availability by a wide margin, so they must accept incorrect specifications for many tasks. Our coverage is lower than theirs, but it is the closest to seed availability on both benchmarks.

Overall, specifications accepted by contrapuntal languages kill the most mutants and are most often correct. Our method rejects more tasks than the baselines, but its coverage is the closest to the upper bound on correct acceptance. This conservatism matters for verification: an incorrect specification can let a buggy program pass verification, whereas a rejected task is simply flagged as unresolved. Contrapuntal languages thus offer the best balance between reliability and coverage.

\vspace{-2mm}

\subsection{RQ2 (Ablation): Impact of Contrapuntality}
\label{sec:rq2}

To assess the contribution of contrapuntality to the effectiveness of our method, we compare our approach with a relaxed version. This variant randomly partitions the sampled seed formalizations into two disjoint subsets of equal size and synthesizes one language independently from each subset, breaking contrapuntality guarantee between two languages. Relative to Algorithm~\ref{alg:synthesis}, it removes the requirement that both languages express every seed, the different-shape constraint in \textsc{Cut}, and the \textsc{FindIso} check and its associated rejection of candidate extensions. We retain the same subsequent formalization and equivalence checking procedure.

Compared with our full approach, the relaxed version yields average relative decreases of 15\% in CR and 34.5\% in Avg.~CR, which support the contribution of contrapuntality: constructing two different languages alone does not provide the same validation reliability as contrapuntal languages.

\subsection{RQ3: Mitigating Correlated Errors}
\label{sec:error-reproduction}

Contrapuntal languages are intended to decorrelate errors: two incorrect formalizations in different languages should be less likely to be incorrect in the same way. We test this by comparing single-language pairs (two specifications generated directly in SMT-LIB) with cross-language pairs (one specification from each contrapuntal language). For each pair type, we measure \emph{error coincidence} (EC): among pairs of incorrect specifications, the fraction that are semantically equivalent.

For each task and model setting, we consider the valid specifications among the 30 direct-generation samples and the valid candidates generated in each contrapuntal language before equivalence-based acceptance. We group candidates into solver-confirmed equivalence classes using the same SMT-based equivalence-checking procedure as in Section~\ref{sec:rq1}. A class is labeled incorrect if a member rejects a reference input--output pair or an evaluated mutant is confirmed to survive its checks. Otherwise, it is labeled correct if a member meets the $0.8$-completeness and soundness criteria used in RQ1. Classes without sufficient evidence for either label are excluded. 

For the single-language comparison, we form unordered pairs of distinct SMT-LIB candidates; for the cross language comparison, we pair one candidate from each contrapuntal language. Pairs with unresolved equivalence checks are excluded from both the numerator and denominator. For task $t$ and pair type $p \in \{\mathrm{single}, \mathrm{cross}\}$, let $D_t^p$ be the number of remaining pairs, $W_t^p$ the number in which both candidates are incorrect, and $E_t^p$ the number in which both belong to the same incorrect equivalence class. T are the tasks with $D_t^p > 0$ for both $p$. We normalize pair counts within each task and then aggregate the resulting proportions across tasks to compute EC:
\[
w_t^p = \frac{W_t^p}{D_t^p},
\qquad
e_t^p = \frac{E_t^p}{D_t^p},
\qquad
\mathrm{EC}^p =
\frac{\sum_{t \in T} e_t^p}{\sum_{t \in T} w_t^p}.
\]

Averaging EC equally across the three model settings, the rate decreases from 25.23\% for single-language pairs to 21.88\% for cross-language pairs on HumanEval+, and from 10.07\% to 5.72\% on LiveCodeBench. Thus, when both candidates are incorrect, cross-language pairs agree on the same incorrect semantics less frequently on average. These results provide empirical support for using contrapuntal representations to decorrelate model errors for more effective validation of autoformalizaed specifications, beyond enforcing differences in AST shape.

\subsection{RQ4: Token Usage}

Table~\ref{tab:token_usage} reports the mean input, output, and total token usage per task under the same 30-response sampling budget used in RQ1.

\begin{table}[h]
\vspace{-3mm}
\captionsetup{aboveskip=3pt, belowskip=3pt}
\caption{Token usage per task on HumanEval+ and LiveCodeBench.}
\label{tab:token_usage}
\centering
\footnotesize
\renewcommand{\arraystretch}{0.85}
\begin{tabular}{llrrrrrr}
\toprule
\multirow{2}{*}{Model} & \multirow{2}{*}{Method} & \multicolumn{3}{c}{HumanEval+} & \multicolumn{3}{c}{LiveCodeBench} \\
\cmidrule(lr){3-5}\cmidrule(lr){6-8}
& & Input & Output & Total & Input & Output & Total \\
\midrule
\multirow{3}{*}{\shortstack[l]{GPT-5.4 mini}}
& Round-trip Veri. & 24,103 & 4,560 & 28,664 & 27,490 & 6,079 & 33,570 \\
& Plurality voting & 32,146 & 6,503 & 38,649 & 39,826 & 8,832 & 48,659 \\
& Contrapuntal langs. & 45,840 & 4,180 & 50,020 & 76,758 & 6,345 & 83,102 \\
\addlinespace
\multirow{3}{*}{\shortstack[l]{Qwen3.6-Flash\\ thinking=off}}
& Round-trip Veri. & 15,396 & 8,149 & 23,545 & 16,658 & 20,070 & 36,728 \\
& Plurality voting & 32,146 & 22,262 & 54,408 & 39,826 & 59,951 & 99,777 \\
& Contrapuntal langs. & 53,375 & 12,514 & 65,889 & 66,124 & 32,848 & 98,972 \\
\addlinespace
\multirow{3}{*}{\shortstack[l]{Qwen3.6-Flash\\ thinking=on}}
& Round-trip Veri. & 15,203 & 3,513 & 18,716 & 18,949 & 15,383 & 34,331 \\
& Plurality voting & 19,968 & 6,050 & 26,019 & 39,826 & 32,454 & 72,280 \\
& Contrapuntal langs. & 46,211 & 4,080 & 50,291 & 75,319 & 22,931 & 98,250 \\
\bottomrule
\end{tabular}
\vspace{-3mm}
\end{table}

All three methods consume more tokens on LCB than on HEP, consistent with its more complex task descriptions and specifications. Round-trip verification consumes fewer tokens than plurality voting because of shorter prompts and condensed task descriptions. Compared with plurality voting, contrapuntal languages consume more input tokens because their prompts include the synthesized language constructs and instructions for using them, also yielding more compact outputs. Their more compact outputs generally do not offset the larger input completely.

On average, contrapuntal languages consume $1.4\times$ as many tokens as plurality voting and $2.5\times$ as many as round-trip verification, using the additional context to induce structurally different representations. This trade-off yields the higher validation performance demonstrated in \Cref{sec:rq1} by addressing correlated errors that additional sampling alone does not remove.

\subsection{RQ5: Analysis of Failure}
\label{failure}

Across the two benchmarks and three model settings, contrapuntal languages' false-positive failures accounted for 38.9\% of all 216 failures. Although we introduce representational diversity between contrapuntal languages, the two languages still share the same seed formalizations and therefore cannot eliminate correlated errors completely. After investigation, most false-positive cases lacked any $0.8$-complete and sound specification among seed formalizations, and among tasks with at least two valid seeds, these cases exhibited approximately 15\% lower normalized equivalence-class entropy than other tasks that lacked such a seed specification, suggesting a more concentrated distribution of incorrect interpretations that may be more difficult to decorrelate even when formalizations are required to use structurally different representations.

False-negative failures accounted for the remaining 61.1\%. These comprised cases with no valid specifications under one or both languages (20.8\%), cases with valid specifications but no matching pairs between the two languages (31.5\%), and failures to synthesize languages satisfying contrapuntality (8.8\%). Invalid specifications primarily resulted from SMT-LIB syntax, type, or symbol errors; failure to use constructs from the synthesized language; and empty outputs or malformed specification structures. The no-matching-pair cases comprised two groups: those containing at least one candidate pair whose equivalence remained unresolved by the solver mostly due to timeout, accounting for 17.6\%, and those in which all candidate pairs were proven non-equivalent, accounting for 13.9\%.

\subsection{Case Study: Ambiguous Task Descriptions}
\label{sec:rq4}

\begin{figure}
    \centering
    \includegraphics[width=\linewidth]{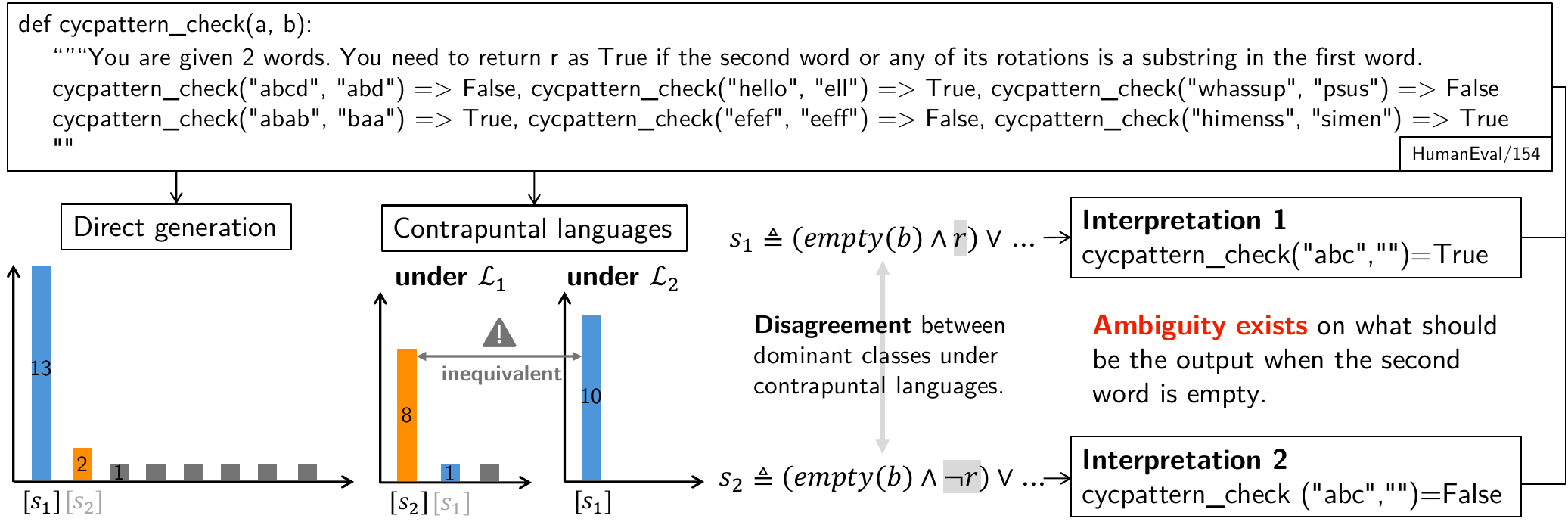}
    \setlength{\abovecaptionskip}{-8pt}
    \caption{Cross-language disagreement exposes different plausible interpretations of HumanEval/154. For an empty second word, $s_1$ requires \texttt{True} and $s_2$ requires \texttt{False}; concentrated distributions under the two languages favor different interpretations.}
    \label{fig:ambi}
    \vspace{-5mm}
\end{figure}

While manually examining changes in candidate distributions across representations, we encountered HumanEval/154, where a reference-consistent interpretation frequent in direct formalization lost its dominance under one of the two languages. Both language-conditioned distributions nevertheless remained concentrated. Inspecting their conflicting dominant specifications revealed an unresolved boundary case in the task description. We examine this case to show how cross-language disagreement can make different plausible interpretations apparent even when each language induces strong agreement. The results below use Qwen3.6-Flash in thinking mode.

HumanEval/154 asks whether a word or any of its rotations is a substring of another, without explicitly addressing an empty second word. Of 30 direct formalizations, 22 were valid and formed semantic classes of sizes $[13,2,1\times7]$. As shown in \Cref{fig:ambi}, the largest class, $s_1$, required \texttt{True} for \texttt{a="abc"} and \texttt{b=""}, matching the benchmark reference. The class of size two, $s_2$, required \texttt{False}. Plurality voting selected specifications representing $s_1$, while the plausible empty-word interpretation appeared in only a small minority of the direct candidates.

Under the contrapuntal languages, the class-size distributions were $[8,1,1]$ in $\mathcal{L}_1$ and $[10]$ in $\mathcal{L}_2$. The previously minor interpretation $s_2$ now accounted for eight of ten candidates in $\mathcal{L}_1$, whereas all ten candidates in $\mathcal{L}_2$ represented $s_1$. Each distribution was more concentrated around its dominant interpretation, but the two dominant classes were not equivalent. Stronger within-language agreement therefore did not settle the empty-word convention; comparing the languages brought another plausible interpretation into focus. This diagnosis is consistent with the SpecFix analysis~\cite{specfix}, in which HumanEval/154 is among the 20 tasks identified as ambiguous by all models in their experiment. The cross-language contrast thus identifies a specific decision for the user.

This case study suggests a complementary role for representational diversity: beyond supporting specification validation, it may help expose ambiguity by making different plausible interpretations explicit. Future work could investigate when such disagreement is indicative of ambiguity and how distinguishing behaviors can be used to formulate targeted clarification questions for users.

\section{Threats to Validity}
\label{sec:threats}

\paragraph{Behavioral evaluation.}
Our completeness and soundness judgments are relative to the tested behaviors, not exhaustive semantic correctness. In particular, killing a mutant establishes only that at least one observed incorrect output is rejected. We apply the same tests and criteria to all methods, providing a consistent basis for comparison within the intended input domain.

\vspace{-1mm}
\paragraph{Test-suite and mutant selection.}
Test reduction may omit inputs that distinguish specifications. Our reduction criteria preserve complementary indicators of test effectiveness, including branch coverage and mutant detection, but these concern the programs used in reduction rather than exhaustive coverage of specification errors. Mutant selection also limits the faults represented; combining real wrong-answer submissions with LLM-generated variants on LiveCodeBench broadens these sources beyond generated mutations alone.

\vspace{-1mm}
\paragraph{Generalizability and interpretation of intent.}
Our evaluation covers two benchmarks of different complexity and three configurations spanning two model families and both thinking and non-thinking modes. Generalization to other settings and specification formalisms requires further evaluation. Reference outputs provide one interpretation of benchmark intent, but ambiguous descriptions may admit others. Our case study shows that cross-language disagreement can surface plausible alternatives, not determine which matches the user's intent. Resolving such ambiguity requires confirmation from the requirement author or user, particularly for real-world requirements that may exhibit more substantial and diverse ambiguities.

\section{Related Work}
\label{sec:related-work}

\paragraph{LLM-based autoformalization and specification generation.}
LLM-based autoformalization has been studied for mathematical statements and software specifications across theorem-prover languages, regular expressions, first-order and temporal logic~\cite{NEURIPS2022_d0c6bc64,hahn2022formalspecificationsnaturallanguage}.
In software engineering, recent work generates postconditions and richer contracts, often using program analysis or verification feedback~\cite{endres2024largelanguagemodelstransform,11029962,10.1007/978-3-031-65630-9_16,richter2025postconditionslargelanguagemodels}.
These works generate specifications in a given target formalism. However, generated specifications are themselves untrusted, and validating whether they capture the intended behavior without an external oracle remains challenging~\cite{lahiri2026intent}. We study oracle-free validation of autoformalized specifications, using two task- specific formal representations as validation views.

\vspace{-1mm}
\paragraph{Consistency-based validation and candidate selection.}
Li et al.~\cite{NEURIPS2024_6034a661} propose \emph{Symbolic Equivalence} (SymEq), which scores candidates by equivalence-class size, and \emph{Semantic Consistency} (SemCo), which scores back-translations by embedding similarity. We adapt SymEq into our plurality voting baseline by accepting specifications in the largest SMT-checked equivalence class. This provides a direct class-based acceptance rule, whereas using SemCo for acceptance would require an additional policy beyond its candidate-ranking scores. Amrollahi et al.~\cite{amrollahi2026faithfulautoformalizationroundtripverification} validate formalizations by checking equivalence after back-translation and re-formalization, and use verification failures to guide subsequent repair. We adopt their complete round-trip verification procedure as our baseline. 
Liu et al.~\cite{liu2026faithfulautoformalizationnaturallanguage} query users to disambiguate candidate assertions when needed, while Mendoza et al.~\cite{mendoza2026automating} use user judgments on distinguishing traces; these oracle-assisted workflows fall outside our oracle-free comparison. Semantic triangulation mitigates correlated errors through consistency across original and transformed programming problems~\cite{dai2026reducinghallucinationsllmgeneratedcode}, but targets code selection; adapting it would require specification-level transformations and consistency relations. We instead preserve task intent while varying its formal representation, validating agreement across two synthesized, structurally distinct languages.

\vspace{-1mm}
\paragraph{Decomposition, reuse, and intermediate representations.}
Several systems improve specification synthesis through decomposition or fragment reuse.
Clause2Inv reuses clauses from candidate invariants~\cite{cao2025clause2inv}, while nl2spec and SynthTL expose structured sub-translations for correction or localized validation~\cite{10.1007/978-3-031-37703-7_18,10918520}.
Req2LTL introduces a hierarchical intermediate representation~\cite{10.1109/ASE63991.2025.00104}, and KELPS studies aligned formalization across theorem-prover languages~\cite{zhang2025kelpsframeworkverifiedmultilanguage}.
We likewise exploit intermediate structure, but synthesize two deliberately different decompositions of the same seeds, combining common seed coverage with structural difference to obtain validation evidence. We do not use these systems as baselines because they address generation, repair, or logic-specific translation rather than the validation setting studied here.

\vspace{-1mm}
\paragraph{Evaluation of generated specifications.}
When a reference formalization is available, candidates can be evaluated against it; for example, BEq+ establishes bidirectional implication using deterministic Lean proofs~\cite{poiroux-etal-2025-reliable}.
Without references, prior work relies on behavioral evidence: Endres et al.~\cite{endres2024largelanguagemodelstransform} evaluate postconditions using trusted tests and buggy mutants, while Lahiri~\cite{10918369} uses symbolic testing with mutated outputs.
FormalBench combines deductive verification with mutation analysis~\cite{le-cong-etal-2025-llms}, while Coins evaluates Rocq specifications on positive and mutation-derived negative cases through concrete proof obligations~\cite{yang2026powerful}.
Without reference formal specifications, we follow this test-based evaluation approach, using positive tests, negative tests, and mutant killing as behavioral approximations of specification quality.

\vspace{-1mm}
\paragraph{SMT-based synthesis and repair.}
Constraint solving has long been used to synthesize programs and repairs. Oracle-guided component-based synthesis encodes the composition of program components as SMT constraints~\cite{10.1145/1806799.1806833}, while DirectFix builds on component-based synthesis and MaxSAT to search for simple program repairs~\cite{2818754.2818811}. Our setting changes the object being synthesized: we use SMT to construct formal languages rather than programs or patches. The encoding searches over cuts of seed formulas and constrains the resulting fragment-tree shapes to obtain an effectively contrapuntal language pair.
\section{Conclusion}

We presented an oracle-free approach to validating autoformalized specifications through \emph{contrapuntal languages}, pairs of task-specific formal languages, each designed to express the intended specification, yet admitting no structure-preserving translation between them. By guiding LLM to formalize the same intent along different representational paths, our approach uses solver-checked agreement as evidence of correctness. Experiments on HumanEval+ and LiveCodeBench show that our approach improves the proportion of accepted specifications that are complete and sound over the strongest baseline on average, supporting deliberately structured representational diversity as a means of making agreement more informative for specification validation.

\begin{acks}
This work was supported by the National Natural Science Foundation of China under Grant No.~W2632069. We thank Haotian Xu, Sijie Liang, and Dimitrios Stamatios Bouras for their feedback.
\end{acks}

\bibliographystyle{ACM-Reference-Format}
\bibliography{ref}

\pagebreak
\appendix

\begin{center}
{\Large\bfseries Supplementary Materials}
\end{center}
\addcontentsline{toc}{section}{Supplementary Materials}

\section{Example of Syntactic Sugar}
\label{app:sugar}

\begin{figure}[b]
\centering
\begin{tikzpicture}[
  every node/.style={inner sep=1.5pt},
  blob/.style={line width=5mm, line cap=round, #1},
  con/.style={fill=#1, rounded corners=2pt, inner sep=2.5pt},
  hole/.style={inner sep=1pt},
]
\newcommand{\blobnode}[2]{\fill[#1] (#2) circle (2.5mm);}
\newcommand{\blobedge}[3]{\draw[blob=#1] (#2.center) -- (#3.center);}
\begin{scope}[shift={(0mm,0mm)}]
  \node[posl] at (0,6mm) {$D_{\kappa_\lor}$};
  \node (o) at (0,0) {$\lor$};
  \node (oe) at (-5mm,-7mm) {$\doteq$};
  \node (or) at (-8.5mm,-14mm) {$r$};
  \node (om) at (-1.5mm,-14mm) {$-1$};
  \node[hole] (oh) at (5mm,-7mm) {$\square_1$};
  \begin{pgfonlayer}{background}
    \blobedge{frPeach}{o}{oe}\blobedge{frPeach}{oe}{or}\blobedge{frPeach}{oe}{om}
  \end{pgfonlayer}
  \draw (o)--(oe) (oe)--(or) (oe)--(om) (o)--(oh);
\end{scope}
\begin{scope}[shift={(27mm,0mm)}]
  \node[posl] at (0,6mm) {$D_{\kappa_{d\land}}$};
  \node (n) at (0,0) {$\land$};
  \node (nd) at (-5mm,-7mm) {$d$};
  \node (nr) at (-8.5mm,-14mm) {$r$};
  \node (na) at (-1.5mm,-14mm) {$a$};
  \node[hole] (nh) at (5mm,-7mm) {$\square_1$};
  \begin{pgfonlayer}{background}
    \blobedge{frBlue}{n}{nd}\blobedge{frBlue}{nd}{nr}\blobedge{frBlue}{nd}{na}
  \end{pgfonlayer}
  \draw (n)--(nd) (nd)--(nr) (nd)--(na) (n)--(nh);
\end{scope}
\begin{scope}[shift={(52mm,0mm)}]
  \node[posl] at (0,6mm) {$D_{\kappa_\land}$};
  \node (c) at (0,0) {$\land$};
  \node[hole] (c1) at (-4.5mm,-7mm) {$\square_1$};
  \node[hole] (c2) at (4.5mm,-7mm) {$\square_2$};
  \begin{pgfonlayer}{background}\blobnode{frGreen}{c}\end{pgfonlayer}
  \draw (c)--(c1) (c)--(c2);
\end{scope}
\begin{scope}[shift={(74mm,0mm)}]
  \node[posl] at (0,6mm) {$D_{\kappa_d}$};
  \node (e) at (0,0) {$d$};
  \node (er) at (-3.5mm,-7mm) {$r$};
  \node (ea) at (3.5mm,-7mm) {$a$};
  \begin{pgfonlayer}{background}\blobedge{frPink}{e}{er}\blobedge{frPink}{e}{ea}\end{pgfonlayer}
  \draw (e)--(er) (e)--(ea);
\end{scope}
\begin{scope}[shift={(94mm,0mm)}]
  \node[posl] at (0,6mm) {$D_{\kappa_\ell}$};
  \node (f) at (0,0) {$\ell$};
  \node (fr) at (-3.5mm,-7mm) {$r$};
  \node (fa) at (3.5mm,-7mm) {$a$};
  \begin{pgfonlayer}{background}\blobedge{frYellow}{f}{fr}\blobedge{frYellow}{f}{fa}\end{pgfonlayer}
  \draw (f)--(fr) (f)--(fa);
\end{scope}
\newcommand{\psinodes}[1]{
  \node (#1v) at (0,0) {$\lor$};
  \node (#1e) at (-16mm,-8mm) {$\doteq$};
  \node (#1er) at (-20mm,-16mm) {$r$};
  \node (#1em) at (-12mm,-16mm) {$-1$};
  \node (#1w) at (16mm,-8mm) {$\land$};
  \node (#1d) at (9mm,-16mm) {$d$};
  \node (#1dr) at (5.5mm,-24mm) {$r$};
  \node (#1da) at (12.5mm,-24mm) {$a$};
  \node (#1l) at (23mm,-16mm) {$\ell$};
  \node (#1lr) at (19.5mm,-24mm) {$r$};
  \node (#1la) at (26.5mm,-24mm) {$a$};
}
\begin{scope}[shift={(0mm,-24mm)}]
  \node at (-6mm,0) {$\chi_1$};
  \node[con=frPeach] (x1) at (8mm,0) {$\kappa_\lor$};
  \node[con=frBlue] (x2) at (8mm,-10mm) {$\kappa_{d\land}$};
  \node[con=frYellow] (x3) at (8mm,-20mm) {$\kappa_\ell$};
  \draw (x1)--(x2) (x2)--(x3);
  \draw[-{Stealth}, thick] (20mm,-10mm) -- node[above] {$D$} (36mm,-10mm);
\end{scope}
\begin{scope}[shift={(66mm,-24mm)}]
  \psinodes{a}
  \node[posl, anchor=west] at ([xshift=1pt]av.east) {$\varepsilon$};
  \node[posl, anchor=east] at ([xshift=-1pt]ae.west) {$1$};
  \node[posl, anchor=east] at ([xshift=-1pt]aer.west) {$11$};
  \node[posl, anchor=west] at ([xshift=1pt]aem.east) {$12$};
  \node[posl, anchor=west] at ([xshift=1pt]aw.east) {$2$};
  \node[posl, anchor=east] at ([xshift=-1pt]ad.west) {$21$};
  \node[posl, anchor=north] at ([yshift=-1pt]adr.south) {$211$};
  \node[posl, anchor=north] at ([yshift=-1pt]ada.south) {$212$};
  \node[posl, anchor=west] at ([xshift=1pt]al.east) {$22$};
  \node[posl, anchor=north] at ([yshift=-1pt]alr.south) {$221$};
  \node[posl, anchor=north] at ([yshift=-1pt]ala.south) {$222$};
  \begin{pgfonlayer}{background}
    \blobedge{frPeach}{av}{ae}\blobedge{frPeach}{ae}{aer}\blobedge{frPeach}{ae}{aem}
    \blobedge{frBlue}{aw}{ad}\blobedge{frBlue}{ad}{adr}\blobedge{frBlue}{ad}{ada}
    \blobedge{frYellow}{al}{alr}\blobedge{frYellow}{al}{ala}
  \end{pgfonlayer}
  \draw (av)--(ae) (ae)--(aer) (ae)--(aem) (aw)--(ad) (ad)--(adr) (ad)--(ada) (al)--(alr) (al)--(ala);
  \draw[cut] (av)--(aw);
  \draw[cut] (aw)--(al);
\end{scope}
\begin{scope}[shift={(0mm,-55mm)}]
  \node at (-6mm,0) {$\chi_2$};
  \node[con=frPeach] (y1) at (8mm,0) {$\kappa_\lor$};
  \node[con=frGreen] (y2) at (8mm,-10mm) {$\kappa_\land$};
  \node[con=frPink] (y3) at (3mm,-20mm) {$\kappa_d$};
  \node[con=frYellow] (y4) at (13mm,-20mm) {$\kappa_\ell$};
  \draw (y1)--(y2) (y2)--(y3) (y2)--(y4);
  \draw[-{Stealth}, thick] (20mm,-10mm) -- node[above] {$D$} (36mm,-10mm);
\end{scope}
\begin{scope}[shift={(66mm,-55mm)}]
  \psinodes{b}
  \begin{pgfonlayer}{background}
    \blobedge{frPeach}{bv}{be}\blobedge{frPeach}{be}{ber}\blobedge{frPeach}{be}{bem}
    \blobnode{frGreen}{bw}
    \blobedge{frPink}{bd}{bdr}\blobedge{frPink}{bd}{bda}
    \blobedge{frYellow}{bl}{blr}\blobedge{frYellow}{bl}{bla}
  \end{pgfonlayer}
  \draw (bv)--(be) (be)--(ber) (be)--(bem) (bd)--(bdr) (bd)--(bda) (bl)--(blr) (bl)--(bla);
  \draw[cut] (bv)--(bw);
  \draw[cut] (bw)--(bd);
  \draw[cut] (bw)--(bl);
\end{scope}
\end{tikzpicture}
\caption{Syntactic sugar on Formula~I of \Cref{fig:comparison}, with $d$ and $\ell$ abbreviating $\mathit{descend\_at}$ and $\mathit{largest}$. Top: the definitions of the five constructs of $\mathcal{L}$, with their non-hole parts colored. Middle and bottom: the two sugarings $\chi_1$ and $\chi_2$ of $\psi_{\mathrm{I}}$ (left) and their common desugaring $\psi_{\mathrm{I}}$ (right), carved into the fragments covered by their constructs (same colors); double strokes mark the cuts between fragments, and gray labels give positions.}
\Description{Top row: five small trees defining constructs, each colored differently, with square holes left uncolored. Below, two rows each show a small tree of colored constructs on the left, an arrow labeled D, and on the right the eleven-node tree of Formula I partitioned into colored regions matching the constructs, with cut marks on the edges between regions.}
\label{fig:sugar-example}
\end{figure}
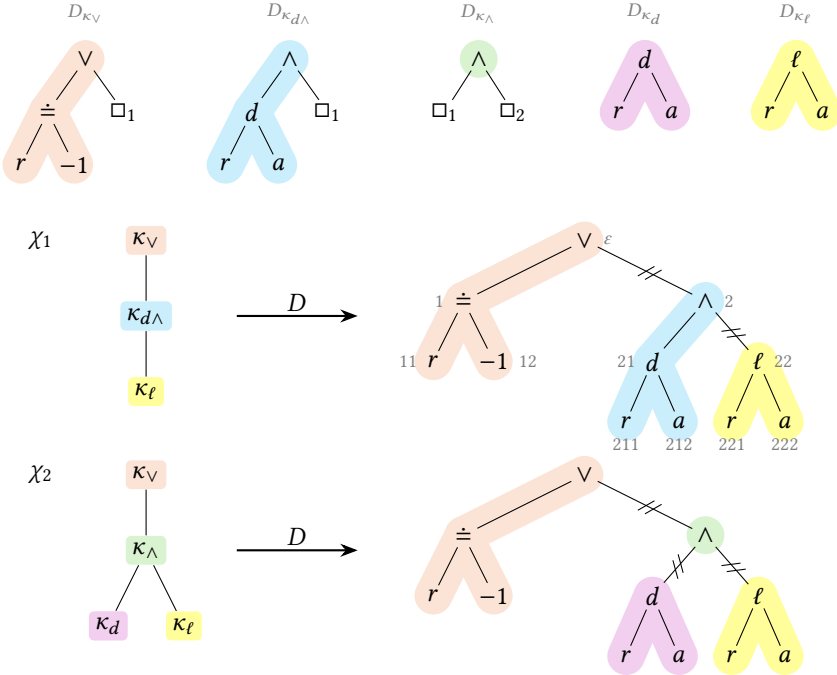

This appendix illustrates the notions of \Cref{sec:background} on Formula~I of \Cref{fig:comparison}. Writing $d$ and $\ell$ for $\mathit{descend\_at}$ and $\mathit{largest}$, the formula is
\[
\psi_{\mathrm{I}} \;=\; r \doteq -1 \,\lor\, (d(r,a) \land \ell(r,a)).
\]

\paragraph*{Trees and positions} $\psi_{\mathrm{I}}$ is a tree with eleven nodes, shown with its positions in \Cref{fig:sugar-example} (middle right). For example, $\psi_{\mathrm{I}}(\varepsilon) = \lor$ is the label of the root, $\psi_{\mathrm{I}}|_2 = d(r,a) \land \ell(r,a)$ is the subtree at position~$2$, and $\psi_{\mathrm{I}}(21) = d$; moreover, $2 \sqsubset 21$, since position~$21$ lies in the subtree at position~$2$.

\paragraph*{A sugared language} Let $\mathcal{L}$ be the sugared language with the five constructs whose definitions are shown at the top of \Cref{fig:sugar-example}:
\[
D_{\kappa_\lor} = r \doteq -1 \lor \square_1, \quad
D_{\kappa_{d\land}} = d(r,a) \land \square_1, \quad
D_{\kappa_\land} = \square_1 \land \square_2, \quad
D_{\kappa_d} = d(r,a), \quad
D_{\kappa_\ell} = \ell(r,a).
\]
The arity of a construct is its number of holes: $\kappa_\land$ is binary, $\kappa_\lor$ and $\kappa_{d\land}$ are unary, and $\kappa_d$ and $\kappa_\ell$ are constants. As \Cref{def:sugar} requires, every definition uses each of its holes and contains at least one other node.

\paragraph*{Desugaring} The $\mathcal{L}$-formula $\chi_1 = \kappa_\lor(\kappa_{d\land}(\kappa_\ell))$ desugars by replacing each construct with its definition, from the leaves up:
\[
D(\chi_1)
= D_{\kappa_\lor}[D_{\kappa_{d\land}}[D_{\kappa_\ell}]]
= D_{\kappa_\lor}[D_{\kappa_{d\land}}[\ell(r,a)]]
= D_{\kappa_\lor}[d(r,a) \land \ell(r,a)]
= \psi_{\mathrm{I}}.
\]
Hence $\chi_1$ desugars to $\psi_{\mathrm{I}}$, and $\chi_1 \equiv \psi_{\mathrm{I}}$. Likewise, $\chi_2 = \kappa_\lor(\kappa_\land(\kappa_d, \kappa_\ell))$ desugars to $\psi_{\mathrm{I}}$.

\paragraph*{Sugarings} In fact, $\Sug_{\mathcal{L}}(\psi_{\mathrm{I}}) = \{\chi_1, \chi_2\}$. The root of a sugaring must be a construct whose definition matches $\psi_{\mathrm{I}}$ at its root, and only $\kappa_\lor$ does; its hole then receives $\psi_{\mathrm{I}}|_2$, which is covered either by $\kappa_{d\land}$, leaving $\ell(r,a)$ for $\kappa_\ell$, or by $\kappa_\land$, leaving $d(r,a)$ and $\ell(r,a)$ for $\kappa_d$ and $\kappa_\ell$. As \Cref{fig:sugar-example} shows, each sugaring carves $\psi_{\mathrm{I}}$ into fragments: every construct of the sugaring covers one fragment, and every edge of the sugaring corresponds to a cut. The two sugarings have different shapes---three nodes versus four---which is what the non-isomorphism requirement of \Cref{sec:contrapuntal} is about: a second language whose only sugaring of $\psi_{\mathrm{I}}$ has four nodes cannot be related to $\chi_1$ by a structure-preserving translation, but may be related to $\chi_2$.

\section{Proof of Theorem~\ref{thm:correctness}}
\label{app:proof}

\paragraph*{Termination} A seed has finitely many positions, hence finitely many cut sets, so there are finitely many extensions for a seed $\psi$ over any given pair of languages. Within one invocation of \textsc{Synth}, the seed $\psi$ and the languages $\mathcal{L}_1, \mathcal{L}_2$ are fixed, and every iteration of the loop either returns or adds to $B$ the extension $\delta$ that \textsc{Cut} has just returned. By the contract of \textsc{Cut}, $\delta \notin B$ before this step, so $B$ grows strictly, and after finitely many iterations \textsc{Cut} returns $\mathsf{none}$ and the loop exits---provided that every recursive call terminates. Each recursive call receives the tail of $S$, so the recursion depth is at most $n$, and termination follows by induction on the length of $S$.

\paragraph*{Adequacy} Suppose that \textsc{Synth} returns a pair $(\mathcal{L}_1^*, \mathcal{L}_2^*)$. It is returned by the base case, reached through a chain of $n$ invocations, the $k$-th of which integrates $\psi_k$ by an extension $(\Delta\mathcal{L}_1^k, \Delta\mathcal{L}_2^k)$ over $(\mathcal{L}_1^{k-1}, \mathcal{L}_2^{k-1})$ and passes $\mathcal{L}_i^k = \mathcal{L}_i^{k-1} \cup \Delta\mathcal{L}_i^k$ to the next, with $\mathcal{L}_i^0 = \emptyset$ and $\mathcal{L}_i^n = \mathcal{L}_i^*$. By the definition of an extension, the fragment tree of the $i$-th cut set of $\psi_k$ is an $\mathcal{L}_i^k$-formula $\chi_i^k$ with $D(\chi_i^k) = \psi_k$; since the cut set satisfies $\mathrm{Bnd}^i$, no cut separates a bound variable occurrence from its quantifier, so $\chi_i^k$ desugars to $\psi_k$. Since the added constructs are fresh, $\mathcal{L}_i^k \subseteq \mathcal{L}_i^*$ and every construct of $\mathcal{L}_i^k$ has the same definition in $\mathcal{L}_i^*$; hence $\chi_i^k$ is also an $\mathcal{L}_i^*$-formula that desugars to $\psi_k$, and $\chi_i^k \in \Sug_{\mathcal{L}_i^*}(\psi_k)$. Clause~(i) of Definition~\ref{def:effective} thus holds for every seed.

\paragraph*{Non-isomorphism} The $n$-th invocation makes its recursive call only after $\textsc{FindIso}(P \cup \{\psi_n\}, \mathcal{L}_1^*, \mathcal{L}_2^*)$ has returned $\mathsf{none}$, where $P \cup \{\psi_n\} = \Psi$. By the contract of \textsc{FindIso}, there is thus no seed $\psi \in \Psi$ with sugarings $\chi_1 \in \Sug_{\mathcal{L}_1^*}(\psi)$ and $\chi_2 \in \Sug_{\mathcal{L}_2^*}(\psi)$ of the same shape. A structure-preserving translation from $\chi_1$ to $\chi_2$ requires $\mathrm{Pos}(\chi_1) = \mathrm{Pos}(\chi_2)$, that is, the same shape; hence none exists, and clause~(ii) of Definition~\ref{def:effective} holds. \qed

\section{Sample Size Setting}
\label{app:hyperparameters}

To choose a practical sampling budget, we examine how the empirical
distribution over specification equivalence classes varies with sample
size. For each task $x$, we measure this distribution using semantic
entropy:
\begin{equation}
    H_N(x)
    = -\sum_{i=1}^{K_N(x)} p_i \log_2 p_i,
    \qquad
    p_i = \frac{|C_i|}{n_{\mathrm{ok}}(x,N)},
    \label{eq:specification_entropy}
\end{equation}
where $K_N(x)$ denotes the number of specification equivalence classes
formed by the successfully generated specifications among the first $N$
sampling attempts, $C_i$ denotes the $i$-th equivalence class, and
$n_{\mathrm{ok}}(x,N)$ is the number of these successful generations.
Failed generations are excluded without replacement.
We vary $N$ from 5 to 50 in increments of 5, using prefixes of the same
50-sample sequence for each task, and report the arithmetic mean of
task-level entropies across the 161 HumanEval+ tasks.

\begin{figure}[t]
    \centering
    \includegraphics[width=0.5\linewidth]{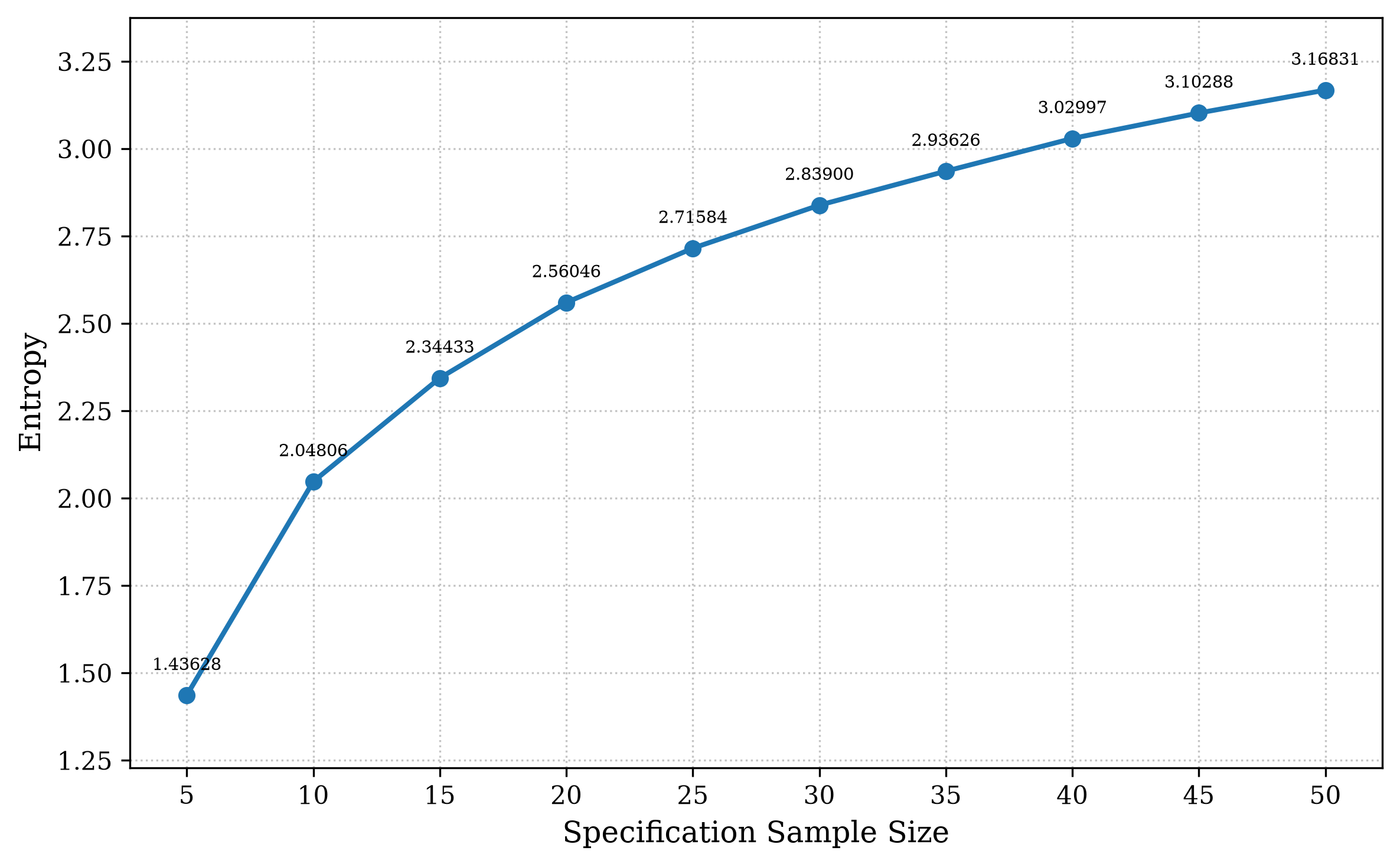}
    \captionsetup{aboveskip=2pt, belowskip=2pt}
    \caption{Mean semantic entropy of specification equivalence-class
    distributions across 161 HumanEval+ tasks as a function of
    specification sample size. Each point averages task-level entropies
    in bits, computed from successful generations among the first $N$
    sampling attempts.}
    \label{fig:specification_entropy}
\vspace{-4mm}
\end{figure}

As shown in Figure~\ref{fig:specification_entropy}, the mean entropy
increases from 1.436 bits at $N=5$ to 2.839 bits at $N=30$, with
progressively smaller gains as the sampling budget grows.
Increasing $N$ further from 30 to 50 raises the mean entropy to
3.168 bits---an additional 0.329 bits, or 11.6\%, despite a 66.7\%
increase in sampling budget.
Although the entropy continues to increase, these diminishing returns
suggest that larger sample sizes provide comparatively modest gains in
measured semantic diversity.
We therefore select $N=30$ as a practical trade-off between representing
specification diversity and limiting sampling and equivalence-analysis
costs.

\section{Full threshold Results for RQ1}
\label{app:threshold}

\begin{figure}[h]
    \vspace{-3mm}
    \centering
    \includegraphics[width=0.95\linewidth]{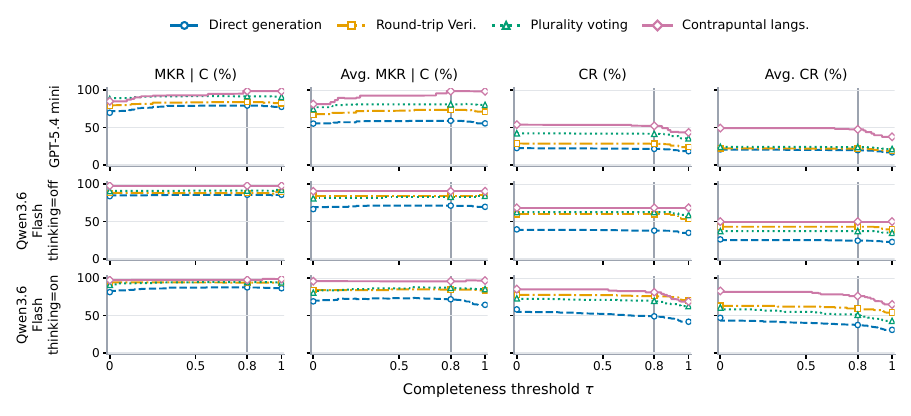}
    \captionsetup{aboveskip=2pt, belowskip=2pt}
    \caption{Threshold sensitivity on HumanEval+. Rows denote models;
    columns denote metrics. Markers show $\tau=0$, $0.8$, and $1$;
    the vertical line marks $0.8$.}
    \label{fig:tau_hep}
    \Description{A three-row, four-column grid comparing four methods on HumanEval+
    across completeness thresholds from zero to one, with rows for model settings
    and columns for sample-level and task-macro joint metrics.}
    \vspace{-3mm}
\end{figure}

\begin{figure}[h]
    \centering
    \includegraphics[width=0.95\linewidth]{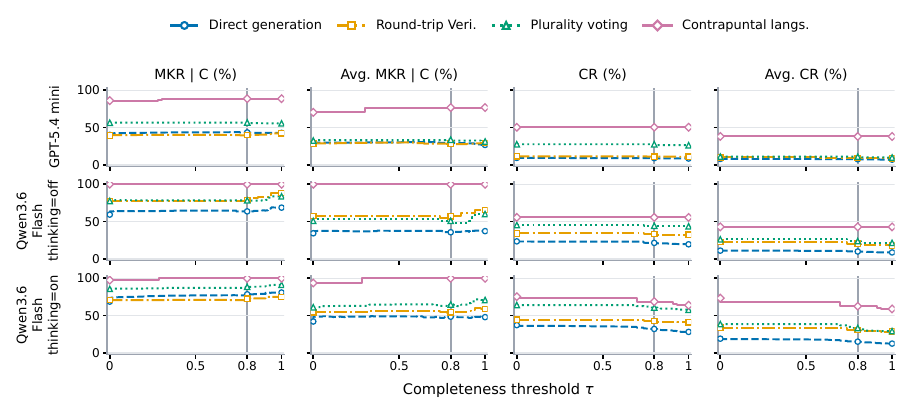}
    \captionsetup{aboveskip=2pt, belowskip=2pt}
    \caption{Threshold sensitivity on LiveCodeBench, using the same layout
    and markers as Figure~\ref{fig:tau_hep}.}
    \label{fig:tau_lcb}
    \Description{A three-row, four-column grid comparing four methods on LiveCodeBench
    across completeness thresholds from zero to one, with rows for model settings
    and columns for sample-level and task-macro joint metrics.}
    \vspace{-3mm}
\end{figure}

This section supplements RQ1 (Section~\ref{sec:rq1}) by examining whether the comparative advantage of contrapuntal languages depends on the completeness threshold. Figures~\ref{fig:tau_hep} and~\ref{fig:tau_lcb} extend the two joint metrics in Table~\ref{tab:hep_result} to the full range $\tau\in[0,1]$. The step curves include every empirical threshold at which the set of $\tau$-complete specifications changes, with the same sample-selection/acceptance and averaging rules as in Table~\ref{tab:hep_result}. Rows correspond to the three model settings. The first two columns report MKR$\mid$C and Avg.~MKR$\mid$C, respectively; the last two report CR and Avg.~CR.

Contrapuntal languages achieve the highest task-level macro-averages, Avg.~MKR$\mid$C and Avg.~CR, throughout $\tau\in[0,1]$ in all six model--benchmark settings. For each sample-level metric, they also lead throughout this range in five of the six settings. There are two exceptions on HumanEval+: with GPT-5.4 mini, plurality voting has a higher MKR$\mid$C at some low thresholds below $0.2$; with Qwen3.6-Flash in thinking mode, round-trip verification has a higher CR at some high thresholds, including $\tau=1$. These results support the overall performance advantage observed in RQ1 across a broad range of thresholds, particularly under task-level macro-averaging.

\end{document}